\RequirePackage{fix-cm}
\documentclass[smallextended]{svjour3}
\smartqed 

\usepackage{natbib}
\usepackage{bm}
\usepackage{diagbox}
\usepackage{lipsum}
\usepackage{lmodern}
\usepackage{soul}
\usepackage{enumitem}
\usepackage{amsmath,amsfonts,amssymb}
\usepackage{mathrsfs}
\usepackage[title]{appendix}
\usepackage{textcomp}
\usepackage{manyfoot}
\usepackage{algorithm}
\usepackage{algorithmicx}
\usepackage{algpseudocode}
\usepackage{listings}
\usepackage{calc}
\usepackage{graphicx}
\usepackage{tabularx}
\usepackage{varwidth}
\usepackage{tasks}
\usepackage{longtable}
\usepackage{supertabular}
\usepackage{float} 
\usepackage{xcolor}
\usepackage{multirow}
\usepackage{multicol}
\usepackage{xspace}
\usepackage[normalem]{ulem}
\usepackage{colortbl}
\usepackage{microtype}
\usepackage{threeparttable,booktabs} 
\usepackage{subcaption}
\usepackage{todonotes}

\newcommand{\revision}[1]{\textcolor{black}{#1}}
\newcommand{\newtext}[1]{\textcolor{black}{#1}}

\renewcommand{\cite}[1]{\citep{#1}}

\usepackage{url}

\def\BibTeX{{\rm B\kern-.05em{\sc i\kern-.025em b}\kern-.08em
    T\kern-.1667em\lower.7ex\hbox{E}\kern-.125emX}}

\usepackage{booktabs, multirow}
\usepackage{pifont}
\usepackage{colortbl} 

\definecolor{verylightgray}{rgb}{0.93, 0.93, 0.93}

\usepackage[skins]{tcolorbox}

\usepackage{tcolorbox}
\tcbuselibrary{skins}
\newenvironment{summarybox}
{\begin{tcolorbox}
[enhanced,arc=0mm,colback=gray!10,frame hidden,overlay unbroken={%
    \draw[thick,black] (interior.north west)--(interior.south west);
},left=2pt,right=0pt,top=0pt,bottom=0pt,before={\vspace{3pt}\noindent},after={\vspace{0pt}}]}
{\end{tcolorbox}}
\newenvironment{summary}
{\vspace{5pt}\noindent\begin{summarybox}}
{\end{summarybox}\vspace{-5pt}}

\newcommand{\code}{Code Generation}
\newcommand{\doc}{Docstring Generation}
\newcommand{\bug}{Bug Fixing}
\newcommand{\test}{Test Generation}

\newcommand{\rqone}{RQ\textsubscript{1}\xspace}
\newcommand{\rqtwo}{RQ\textsubscript{2}\xspace}
\newcommand{\rqthree}{RQ\textsubscript{3}\xspace}
\newcommand{\rqfour}{RQ\textsubscript{4}\xspace}

\begin{document}



\title{The Cost of AI-Assisted Coding: An Extensive Investigation of Energy vs. Accuracy in Language Models}

\titlerunning{The Cost of AI-Assisted Coding}




\author{Negar Alizadeh \and Boris Belchev \and \\ Nishant Saurabh \and Patricia Kelbert \and Fernando Castor}

\institute{
Negar Alizadeh \at
\textit{Utrecht University}\\
Utrecht, The Netherlands \\
\email{n.s.alizadeh@uu.nl}
\and
Boris Belchev \at
\textit{University of Twente}\\
Enschede, The Netherlands \\
\email{b.belchev@student.utwente.nl}
\and
Nishant Saurabh
\at
\textit{Utrecht University}\\
Utrecht, The Netherlands \\
\email{n.saurabh@uu.nl}
\and
Patricia Kelbert
\at
\textit{Fraunhofer IESE}\\
Kaiserslautern, Germany \\
\email{patricia.kelbert@iese.fraunhofer.de}
\and
Fernando Castor
\at
\textit{University of Twente}\\
Enschede, The Netherlands \\
\email{f.castor@utwente.nl}
\and
}

\date{}
\maketitle
\noindent\textcolor{red}{\textbf{Paper under review.}}
\abstract
Developers increasingly rely on generative AI-based coding assistants such as ChatGPT and GitHub Copilot in their workflows.
While many such tools are accessible via remote APIs, data privacy and security concerns drive client organizations toward locally deployed language models.
This study examines the accuracy-energy trade-off in local LLM deployment, aiming to provide valuable insights to help developers make informed decisions when selecting a language model. We evaluated 26 LLM families (including both Mixture of Experts and dense architectures) across common software development tasks on two hardware configurations: a commodity GPU and a high-performance AI-specific GPU. As local LLM deployment demands powerful infrastructure investment, which might not be affordable for everyone, we consider both full-precision and quantized variants.
Our results demonstrate that larger models with higher energy requirements do not consistently yield proportional accuracy gains. Moreover, quantized large models typically outperform full-precision medium-sized models in both efficiency and accuracy. We also find that no single model excels across all software development task types. \textcolor{black}{Finally, the number of active parameters, output length, and quantization level jointly explain over 90\% of the variance in inference energy consumption across all tasks. The results confirm that energy consumption scales with the interaction between output length and model size, consistent with the expected $O(np)$ behavior, while prompt size has a negligible impact.}


\section{Introduction}
\label{sec:intro}
Generative large language models (LLMs) are now widely used in everyday life and across diverse industries, including healthcare, finance, media, and entertainment. According to a recent report by OpenAI~\cite{openai2025chatgptusage}, since the public release of ChatGPT in November 2022, the platform reached more than 100 million logged-in weekly active users within its first year, nearly 350 million after two years, and by July 2025, it had surpassed 700 million total active users, representing almost 10\% of the global adult population. The report also highlights steady growth in work-related usage, but an even faster rise in non-work-related interactions, which increased from 53\% to over 70\% of total messages.
Beyond general-purpose chat-based LLMs, many task-specific models, such as Google Gemini, Claude Code, and GitHub Copilot, have been integrated into IDEs and are now used by both individual developers and organizations. The 2025 Stack Overflow Developer Survey found that 84\% of respondents are already using or plan to adopt AI tools in their workflows~\cite{stackoverflow2025survey}. Similarly, a GitHub study reported that adoption of their AI pair programmer enables developers to code up to 55\% faster, while 85\% of users indicated that it improved their confidence in code quality ~\cite{github2025copilotimpact}.
This high level of adoption underscores the AI assistant's role as an essential tool that enhances productivity, streamlines the coding process, and increases efficiency by providing real-time debugging assistance and automating routine tasks.

Although third-party services are widely used for these purposes, with ChatGPT and GitHub Copilot being the most popular \cite{stack2024aihelp}, reliance on third-party APIs raises concerns for many companies regarding security, data privacy, and subscription costs. Thus, there is a growing interest in local solutions, with many individuals and organizations seeking to set up their own AI coding assistant using open-weight, often smaller language models, but still comparable to publicly available alternatives in specific tasks.
One approach for deploying LLMs locally is to utilize a flagship GPU with sufficient memory optimized for Deep Learning (DL) applications and to set up an open-weight LLM on it. However, since these GPUs are not affordable for everyone, an alternative solution could be using compressed and quantized models that can run on smaller GPUs or even large CPUs.

LLMs can simplify many tasks. Notwithstanding, this convenience comes with significant costs. Their power consumption is a major concern, leading not only to substantial financial expenses but also to high carbon emissions~\cite{strubell2019energy,lacoste2019quantifying}. While we cannot slow down the expansion of DL in daily life, we can raise awareness about its energy consumption and encourage informed decision-making before deploying models.
\\
Competition among AI companies to deliver more capable LLMs has led to the emergence of models that inherently require substantial computational resources and energy. Prior work has examined the energy consumption of language models~\cite{luccioni2022estimating,Patternson:2022:CFM}, but most studies have concentrated on the training phase, with comparatively little attention given to inference, even though inference can account for a substantial share of long-term energy use~\cite{Patternson:2022:CFM}. 
Among the limited studies addressing inference energy consumption~\cite{zschache2025comparing, adamska2025green,fernandez-etal-2025-energy, poddar-etal-2025-towards}, the focus has been almost exclusively on natural language tasks. 
\\
\textcolor{black}{From an AI perspective, many software engineering (SE) tasks can be viewed as sequence-to-sequence problems similar to traditional natural language tasks. However, the research literature considers them as a separate class because they add additional constraints and requirements that are not captured by typical NLP benchmarks~\cite{gao2025systematic, hou2024large}. 
Programming languages impose strict syntactic and semantic rules, and outputs are evaluated based on formal correctness criteria such as compilation, test passing, and semantic preservation, unlike NLP tasks, which often rely on fluency or similarity~\cite{pujar2025code}. Moreover, errors in generated code can lead to severe consequences, such as compilation errors and security vulnerabilities. Therefore, studying energy consumption in SE tasks is only meaningful when considered alongside correctness, as inefficient but correct models and efficient but incorrect models have different practical value in software development settings. In this sense, LLMs in software development tasks differ fundamentally from natural language and therefore require dedicated analysis.
\\
Despite this, recent evaluations of code-specific LLMs have largely focused on accuracy, while their energy footprint has often been overlooked~\cite{zheng2023survey, jiang2024survey}. This work is the first to systematically examine the energy implications of LLMs in coding tasks.
}
\par
In this paper, we evaluate open-weight LLMs in terms of accuracy and energy consumption on two real-world scenarios: the first one utilizes a powerful AI-specific GPU on a cluster, and the second one has direct access to a commodity, end-user GPU. To enable deployment on smaller hardware, we considered both compressed and quantized versions of the LLMs, alongside their full-precision counterparts. We specifically focus on four typical software development tasks, namely Code Generation, Bug Fixing, Docstring Generation, and Test Case Generation, \textcolor{black}{as they are identified among the top AI-assisted programming-related activities in recent surveys~\cite{stackoverflow2025survey} and studies~\cite{watson2022systematic}.}
In particular, we address the following research questions:
\\
\begin{description}
\item \textbf{\rqone} How does the energy consumption of models vary across typical software-related tasks?
\\
\textcolor{black}{For this research question, we quantify and compare the total energy consumption of 78 open-weight model variants (26 models in 3 quantization levels) across four distinct software development activities to identify which tasks impose the highest energy footprint.}
\item \textbf{\rqtwo} How does the performance of different models vary in terms of energy efficiency and accuracy when executing the same task? Is there a trade-off between energy efficiency and accuracy?
\\
\textcolor{black}{This research question examines whether higher accuracy comes at the cost of higher energy use, using a Pareto frontier to identify the most efficient model configurations.}
\item \textbf{\rqthree} Are there specific model characteristics (e.g., feed-forward network dimension, number of transformer blocks) that consistently impact energy?
\\
\textcolor{black}{This research question focuses on analyzing how specific design elements, such as active parameter count, attention heads, and quantization bit-width, statistically correlate with a model's energy efficiency (tokens per joule).}
\item \textbf{\rqfour} What are the performance differences between general-purpose models and models specifically optimized for code-centric tasks?
\\
\textcolor{black}{The last research question aims to evaluate whether code-specific models achieve better energy–accuracy trade-offs compared to general-purpose models when performing the same software engineering tasks.}
\\
\end{description}

\subsection{\textcolor{black}{\textbf{Extension over the Conference Paper}}
}
\label{sec:extension}
\textcolor{black}{
This study expands upon our earlier work published at the MSR 2025 conference~\cite{alizadeh2025language}. In this extended study, we broaden our scope by including more recent language models (e.g., Qwen models), as well as models that incorporate efficiency-oriented techniques which align directly with the goals of our work. These efficient models include Mixture-of-Experts (MoE) models (e.g., DeepSeek-Coder-V2 and Granite3-moe) and models with flexible parameter technology (e.g., Gemma3n). For compatibility with these newly released models, we upgraded Ollama's inference backend. To maintain fair comparisons, we re-executed all experiments, both original and new models, under the updated infrastructure. This resulted in shifts in the original numerical results, so we recalculated correlations in Section~\ref{sec:f3} and revised the interpretation of findings in Section~\ref{sec:res}.
\\
Furthermore, as presented in Section~\ref{sec:regression}, we conducted a regression analysis to quantify the relationships between inference energy consumption and model characteristics, as well as workload factors (input and output token counts). In Section~\ref{sec:cuda} we deepened our investigation by profiling CUDA activities at the kernel level during inference across different tasks to better understand how model operations are mapped to GPU kernels under different tasks and quantization levels within Ollama's backend.
\\
Finally, we expanded the related work section to incorporate recent studies published since the conference version. The replication package is also updated accordingly~\cite{Alizadeh2026_replication}.
\\
In summary, the extended version makes the following contributions:
\begin{itemize}
\item We provide a large-scale empirical evaluation of the energy–accuracy trade-off for 78 variants from 26 language-model families, including both dense and MoE categories.
\item We quantify the drivers of inference energy through linear regression. Active parameter count, output token count, quantization, and their interactions explain 89–94\% of the variance across tasks; in particular, the interaction between output tokens and active parameters is consistently significant.
\item Through CUDA kernel-level profiling, we show that matrix-vector multiplication accounts for more than 64\% of total execution time, identifying the main computational target for improving inference efficiency.
\end{itemize}
}

\section{Language Models}
\label{sec:back_related}
Language models are machine learning models based on the transformer~\cite{Vaswani:2017:AIA} architecture whose main task is, given a sequence of tokens, to predict which tokens are more likely to follow. A language model is composed of multiple transformer blocks, each typically consisting of self-attention mechanisms followed by feed-forward neural networks (FNN).
They can be categorized into three architectural categories: encoder-only, decoder-only, and encoder-decoder models~\cite{jiang2024survey}. Among these, decoder-only models, are optimized for generating text, making them ideal for content creation~\cite{wang2024history}. Code-specific language models, in particular, are pre-trained on extensive unlabeled code corpora, whereas general-purpose (or ``non-coding'') models are trained on a mixture of code and text data \cite{jiang2024survey}. In the remainder of this section, we examine some of the main characteristics of a language model and also briefly introduce the concept of quantized models.

\noindent
\textbf{Attention heads} enable models to process input sequences in parallel. Each attention head can focus on different relationships and patterns in the data.

\noindent
\textbf{Embedding Length} is the dimensionality of the vector space in which tokens are represented. Each token is mapped to a vector of this length in a way that similar tokens are close to each other in the embedding vector space.

\noindent
\textbf{Feed-Forward Length} refers to the size of the intermediate layer in the FNN within each transformer block. The feed-forward length determines the capacity of this part of the network to comprehend complex interactions between tokens.
\noindent
\textbf{A Mixture of Experts (MoE)} architecture replaces standard feed-forward layers in a transformer with many specialized sub-networks, called experts. A separate router network decides which experts to activate per token. This means that although the model may contain a large number of parameters overall, only a small fraction is actually used during each inference. MoE models provide large capacity without proportional compute cost~\cite{grootendorst2024visual}.

\noindent
\textbf{Quantization} is where a model’s weights or activations are represented in a lower-precision format (e.g., 8-bit integers) instead of the original high-precision floating-point format~\cite{jin2024comprehensive}. \textbf{Quantized models} undergo the process of reducing the bit-width of their tensors, which results in a decreased memory footprint and lower computational demands while maintaining acceptable accuracy~\cite{gong2024survey}. This makes quantized models especially valuable for deployment in resource-constrained environments, such as mobile devices or embedded systems, where computational power and memory are limited.
Quantization methods can be broadly grouped into two main categories, quantization-aware training and post-training quantization. The former integrates the quantization process within the training phase and the latter applies quantization after the model has been trained~\cite{jin2024comprehensive}. In this study, as we focus on the inference of LLMs, we specifically examine post-training quantization. Preliminary evidence~\cite{rajputbenchmarking} suggests that, among the multiple quantization formats, GGUF (GPT-Generated Unified Format) is the most energy-efficient format currently available. Therefore, we utilized GGUF to compare the quantized versions of the models under examination.

\section{Methodology}
\label{sec:method}
\begin{figure}[tb]
\centerline{
\includegraphics[width=0.8\linewidth]
{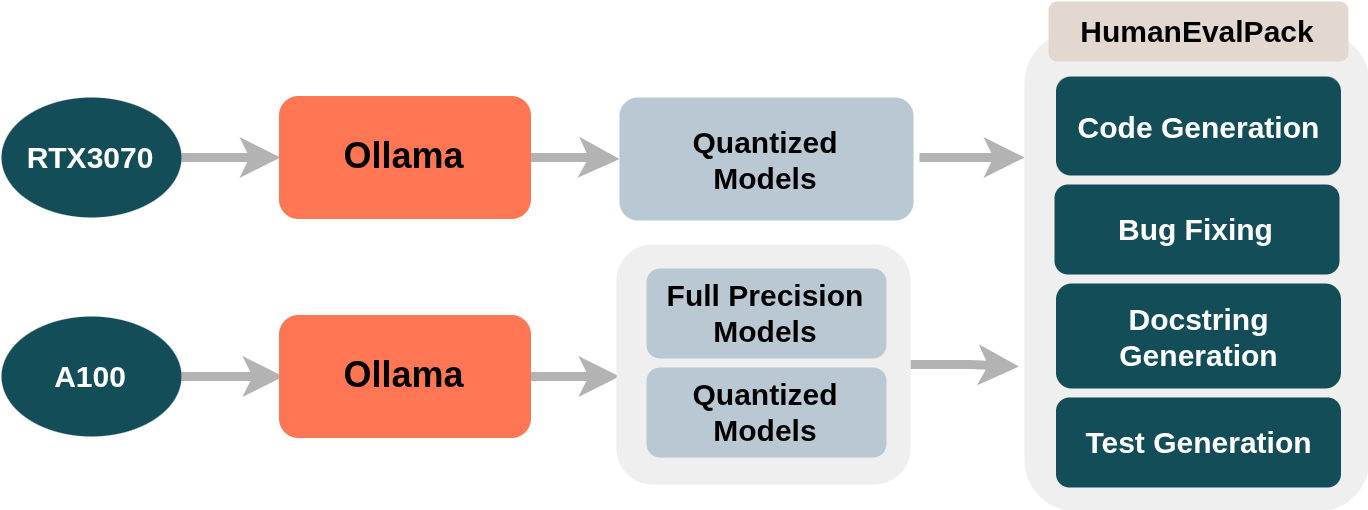}}
\caption{Schematic representation of the study. The circles represent the models of the NVIDIA GPUs available on the two machines where the experiments were run. The experiments involved quantized and full-precision models. We measured the energy footprint of these models when performing four different tasks, using the workload provided by the HumanEvalPack benchmark.}
\label{fig1}
\end{figure}

In this study, we examine the energy consumption of open-weight LLMs when supporting software development tasks, from generating code and docstrings to testing and fixing bugs. Figure~\ref{fig1} presents a high-level overview of how we conducted this study. This section outlines the language models we analyzed, the experimental setup, and the methodology used to conduct our experiments.

\subsection{\textbf{Model Selection}}
We begin our exploration by gathering a list of top LLMs based on the following criteria: \textcolor{black}{popularity (based on Hugging Face downloads), creator reputability, local availability, and compatibility with the GGUF format, which are detailed below. These criteria are applied consistently across all models, and models were included only if they met all of these criteria.}

\textbf{Popularity:} Popularity is measured by the number of downloads of the model sourced from HuggingFace \cite{huggingface_website}. This will ensure that the results apply to the most commonly used models in real-world scenarios. 
All GGUF versions of the models were downloaded from Ollama's library \cite{ollama_library}. \textcolor{black}{Ollama is the runtime framework we used to execute GGUF models locally; it is described in more detail in Section~\ref{sec:hardware}.} Only pages that explicitly referenced the model's original creators' pages were included, excluding those modified by third parties.

\textbf{Reputability of creator:} While this criterion can be arguably subjective it is still crucial. Companies and research teams that have substantial funding and support from other entities in the field are more likely to develop further and support their models. They are assumed to have a larger infrastructure and more research capabilities as they strive for competitiveness with other key players in the field.
Choosing models according to this criterion increases the value of the results as small or medium-sized companies would be more interested in deploying their models.

\textbf{Availability \textcolor{black}{and GGUF compatibility:}} Models that are not publicly available for downloading and setting up locally are automatically excluded from this study, such as GPT-3 and GPT-4. Furthermore, because Ollama supports only GGUF models, all models need to be compatible with conversion to the GGUF format\cite{ggml_gguf}. For example, the Phi3:7b architecture was not supported and could not be included. After failing to find it in the Ollama library, we attempted to convert the original model to GGUF using llama.cpp library \cite{llama_cpp_repo} but encountered an \textit{Unknown architecture} error.
 
As this research focuses on integrating LLMs in software development, we have limited our scope to include both general-purpose (non-coding) models and those fine-tuned for code (coding models). 
We chose instruction-tuned models (additionally trained and polished to understand instructions) over base models because, as shown by the leaderboard \cite{evalplus_leaderboard} maintained by Liu et al. \cite{liu2024your}, instruction-tuned models consistently outperformed their base counterparts in terms of accuracy.
Due to hardware limitations, we also narrowed down our selections to models with no more than 20 billion parameters. The memory constraints of our desktop GPU only allow for quantization levels up to 5 bits. However, \cite{jin2024comprehensive} shows that reducing the bit-width below 3 or 2 significantly degrades model performance. Therefore, we limited our quantized collection to 4 and 5 bit-width for medium-size models (6-8B parameter), and 8 bit-width for small models (2-4B parameter).

We collected all full-precision, i.e., non-quantized, models and their quantized versions in GGUF format from the Ollama library, preserving the original naming convention. Each model name includes the model family and its parameter size; for example, codellama:13b represents the CodeLLaMA model with 13 billion parameters. The suffixes indicate the precision or quantization level used: fp16 denotes full-precision 16-bit floating-point representation, and -q followed by a number specifies the quantized bit width (e.g., q8 represents an 8-bit quantized version of the model).
Two models (Granite 3.3:8b and Granite 3.3:2b) were unavailable on Ollama in their full-precision and Q8 versions, so we obtained their GGUF files directly from the official IBM Granite repository on Hugging Face~\cite{ibm_granite_8b_2024},\cite{ibm_granite_2b_2024}.

\subsection{\textbf{Hardware, Frameworks, and Libraries}}
\label{sec:hardware}
Table~\ref{tab:hardware_setup} presents a summary of the configuration of the two machines employed in this study. We identify each machine in terms of its GPU. Resource-demanding experiments, including full-precision models and their quantized counterparts, were conducted on a machine equipped with an NVIDIA A100 with 80GB memory, accessed via SSH. The jobs were managed using a job scheduling system (Slurm), as the machine is part of a cluster. 
On the other hand, experiments with small quantized models were performed on a laptop equipped with an NVIDIA GeForce RTX 3070 GPU.
To ensure that all the experiments are running on an isolated GPU, on the cluster system, we pass the \textit{exclusive} flag to the job scheduling system, and on the laptop, we stopped GUI and switched to command-line mode. Each experiment was run three times to reduce the impact of transient variations. \textcolor{black}{The reported values are averaged across repeated runs.}

\begin{table}[h]\centering
\caption{Hardware setup}\label{tab:hardware_setup}
\scriptsize
\begin{tabular}{lp{6.5cm}}
\toprule
\multirow{3}{*}{RTX 3070} & \textbf{CPU:} Intel Core i7-11850H - 2.5GHz | Mem: 31GB | Cache: 24MB | Governer: Powersave\\
& \textbf{GPU:} NVIDIA GeForce RTX3070 | Mem: 8GB | PowerMizer: Adaptive | TGP: 220W \\
& \textbf{OS:} Ubuntu 20.04 (64-bit) \\
\midrule
\multirow{3}{*}{A100} & \textbf{CPU:} 2 \ding{53} AMD 7313 - 3GHz | Mem: 1TB | Cache: 32MB | Governer: Performance \\
& \textbf{GPU:} NVIDIA A100 PCIe | Mem: 80GB | PowerMizer: High Performance \\
& \textbf{OS:} AlmaLinux 8.10 (64-bit) \\
\bottomrule
\end{tabular}
\end{table}

\textbf{Running an LLM} locally requires a compatible runtime framework that can handle the model's execution (inference) on various devices. This ensures the model performs efficiently whether it's deployed on desktops, laptops, or another environment. 
For quantized models in the GGUF format, several open-source frameworks have emerged, including PrivateGPT, Ollama, GPT4All, and llamafile \cite{langchain_local_llms}. Among these options, we decided to use the most popular one at the moment, Ollama (v. 0.11.7 with its Python package v. 0.3.2), based on its number of stars on GitHub, and the ease with which it enables us to switch between models. 
\\
\par
\textbf{For measuring energy} and utilization of GPU as in previous studies\cite{alizadeh2024green, argerich2024measuring, castor2024estimating}, we employed the Python binding of the NVIDIA Management Library~\cite{nvidia:2024} (pyNVML \cite{pynvml}) to record the GPU power. The profiling rate is set to 10Hz (one sample every 100ms) to avoid the observed overhead for higher rates \cite{castor2024estimating}.
\textcolor{black}{The energy usage is then determined as the product of execution time and the arithmetic mean of GPU power, computed over all samples collected during runtime.}
As for the energy consumption of the CPU we used a Python library designed for interacting with Intel's RAPL\cite{khan2018rapl} interface named pyRAPL \cite{pyrapl}.
To measure the energy usage of the hardware and operating system when idle, we utilized pyRAPL in the Ubuntu terminal, while only the basic operating system services were running on the machine. 
The sampling rate was 1Hz \cite{intel_rapl} with a total of 3400 observations collected, ensuring a 98\% confidence level that the real power value lies within ±2\% of the observed values. The mean power draw of the CPU in idle mode was 1.92W.
Similarly, for measuring the idle power of GPUs, we executed the nvidia-smi \cite{nvidia_smi} command for 340 seconds to collect 3400 samples at 100ms intervals. The exact command can be found in the online Appendix~\cite{Alizadeh2026_replication}.
The recorded idle power values for RTX3070 and A100 were 9.92W and 46.82W, respectively.

\subsection{\textbf{Dataset}}
\textcolor{black}{In studies assessing the coding proficiency of LLMs, HumanEval\cite{chen2021evaluating} is the most widely-used benchmark \cite{li2023starcoder, lozhkov2024starcoder, roziere2023code, chen2021evaluating, muennighoff2023octopack, du2024mercury} and has become the de facto standard for this purpose~\cite{jiang2024survey}.}
It consists of 164 hand-written Python programming problems and each problem comes with its own set of test cases. 
In this study, we employed an extended version of HumanEval called HumanEvalPack~\cite{muennighoff2023octopack}. 
It includes more rigorous test cases and encompasses all of our desired coding tasks: Code Repair (Bug Fixing), Code Explanation (Generating Docstring), Code Synthesis (Generating Code), and Test Assertion Generation covering six programming languages. We evaluate all models on the Python subset of HumanEvalPack. To ensure that each problem, across all four tasks, is accompanied by an explicit example of input-output pairs in the correct format, we added two new columns, one containing a sample test case assertion and the other an example input-output pair for a docstring. These examples were primarily extracted from the provided fields, with a few cases manually added, and are used in the prompt as the function start for the respective task (online Appendix A~\cite{Alizadeh2026_replication}).

\subsection{\textbf{Experimental Procedure}}

\textbf{Hyperparameters.} 
In our preliminary experiments, we observed that hyperparameter changes can impact energy consumption. To eliminate other confounds, we fixed the hyperparameters for all experiments. 
Following recommendations from prior studies \cite{muennighoff2023octopack, li2023starcoder, lozhkov2024starcoder, roziere2023code, samsi2023words} which employed HumanEval as a coding benchmark, we set the top-p value to 0.95 and the temperature to 0.1 across all experiments to ensure a fair comparison. Previous work~\cite{chen2021evaluating, roziere2023code} has shown that increasing the temperature adversely affects the pass@1 metric, with 0.1 identified as the optimal value.
Finally, to ensure the reproducibility of our experiments, all the results are generated using a fixed seed number.
\\
\par
\textbf{Prompts}
To ensure a fair comparison and consistency with previous studies in which pre-trained code LLMs are evaluated, we did not optimize prompts but used the same format across all models to distinguish question and answer\cite{muennighoff2023octopack, wei2024magicoder, alpaca}. Each prompt includes both the instruction and the function start. Since LLMs are trained on a next-token prediction objective, including the function start helps the model complete it directly without searching within the output~\cite{muennighoff2023octopack}. For each task, we adjusted the information in the prompts accordingly. An example of the prompts is available in the online Appendix A~\cite{Alizadeh2026_replication}.
\\
\par
\textbf{Evaluation}
To assess the correctness of the outputs generated by LLMs, we employed the mean pass@k (mean success rate) evaluation metric, as defined in the Codex evaluation set~\cite{chen2021evaluating}. 
It considers a problem solved if any of the k-generated solutions pass all test cases. We focus on pass@1, which shows the probability of the model solving a problem in one try.
Specifically, for \code~tasks, the correctness of the generated code is determined by whether it passes or not, all the hand-written test assertions provided in the dataset. If the code satisfies all test assertions, it is counted as a correct answer. \textcolor{black}{It is worth noting that, according to the terminology of the Automated Program Repair literature, such solutions should be interpreted as \textit{plausible} rather than definitively correct~\cite{Martinez2025Sustainability}.}
\\
We followed a similar approach for evaluating \bug~tasks, where the correctness of the fixed code was measured by its ability to pass the corresponding test assertions. To assess the correctness of the LLM-generated test assertions, we also followed a similar approach. However, in this case, we executed the canonical solution alongside the generated test assertions. We classified a test as incorrect if the canonical solution fails to pass the assertions, either due to incorrect assertions or syntax errors \cite{li2024large}. To assess the thoroughness of the generated tests, we combined the canonical solution with the generated test assertions in a Python file and measured coverage using the Python code coverage analysis tool, coverage.py \cite{BatchelderCoveragepyThecode}.
To evaluate the correctness of the generated docstrings, we used another LLM as a judge to generate code based on the provided docstring. The generated code was then evaluated using the pass@1 metric. We explored three different options for selecting the judge LLM: (1) feeding the generated docstring back into the model that originally generated it~\cite{muennighoff2023octopack}, (2) selecting the best-performing publicly-available, locally executable LLM among the models under evaluation, and (3) utilizing one of the best proprietary LLMs available, more specifically GPT-4o-mini. Based on our investigation, to avoid potential bias, we decided to proceed with the third option.

\section{Results}
\label{sec:res}
In this section, \textcolor{black}{we present the findings of this study in the same order as the research questions, with each subsection titled to highlight the corresponding key finding.} For the sake of reporting and analyzing results, throughout the section, we treat different quantization levels and numbers of parameters for the same language model family as different models. Models are grouped into families based on the organization developing them, their architecture, and their parameter count. For example, we refer to Gemma:2b-q8 and Gemma:2b-q4 as different models within the same family. However, Gemma:7b-q8 belongs to a different family. This makes sense as they are made available separately, have different memory requirements, and often different architectures.


\subsection{
\textbf{Models exhibit different energy consumption for different software development tasks, and different models vary widely in how they perform the same task.}}\label{sec:f1}
\begin{figure}[ht]
  \centering
  \begin{subfigure}[b]{\textwidth}
    \includegraphics[width=\textwidth]{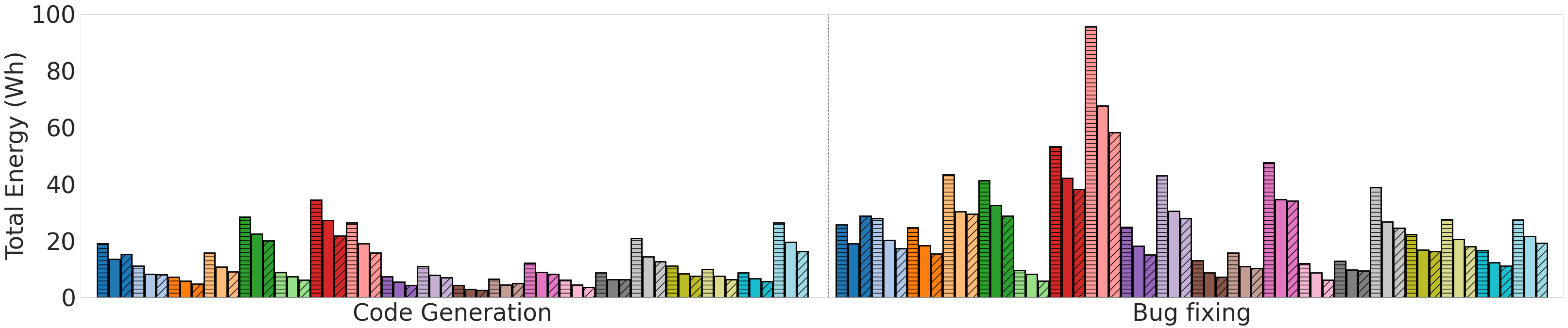}
  \end{subfigure}

  \vspace{0.3cm}

  \begin{subfigure}[b]{\textwidth}
    \includegraphics[width=\textwidth]{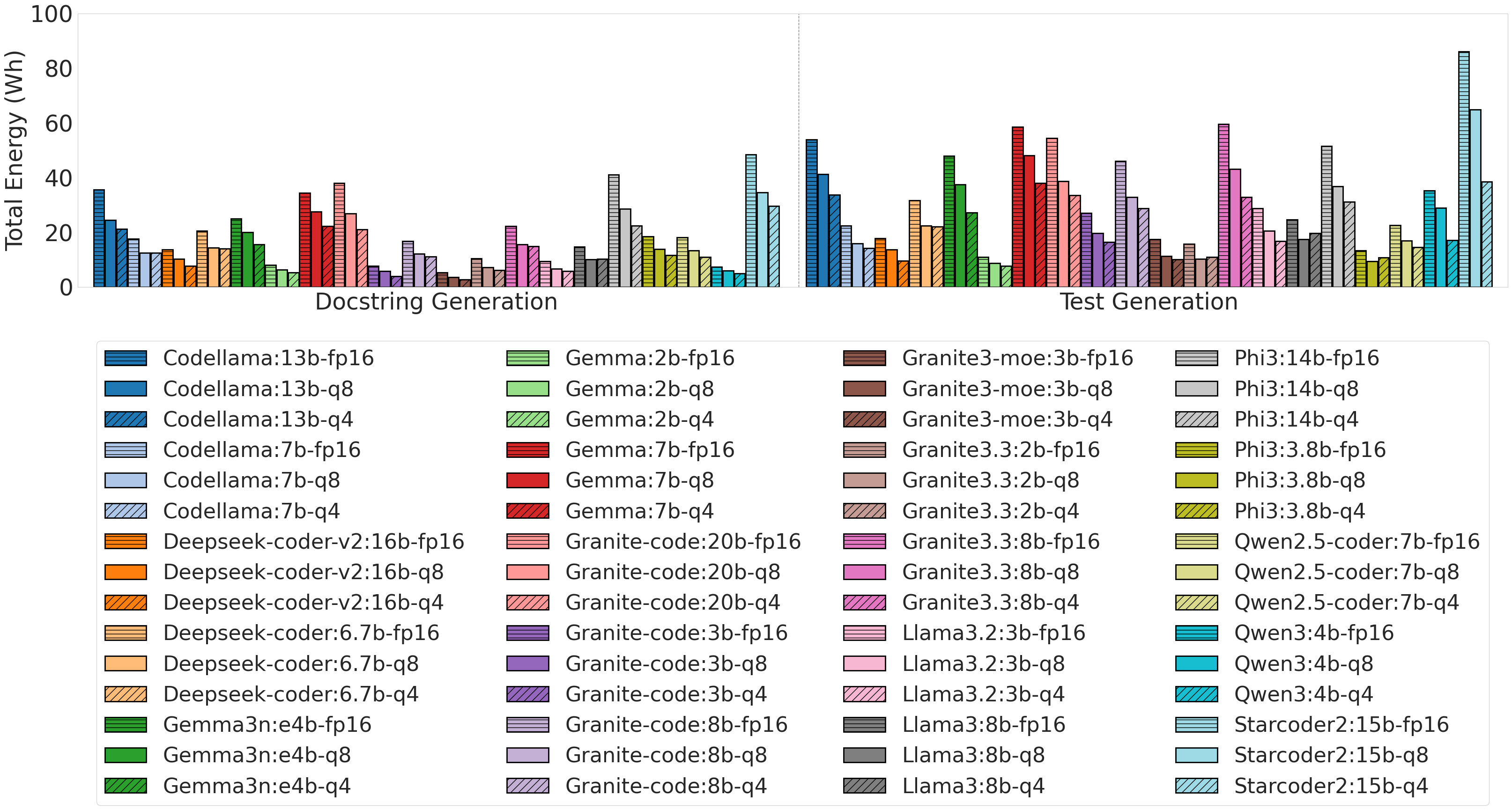}
  \end{subfigure}

  \caption{Total energy (Wh) consumed by the A100 GPU when performing each task. Model names are presented according to the pattern \textit{family-Q}, where \textit{family} includes the family name (Gemma, Phi3, etc.) and the number of parameters, in billions (7b, 3.8b, etc.), and \textit{Q} is the quantization level (4 bit, 8 bit, or full precision 16 bits)}
\label{fig:total_energy_a100}
\end{figure}

\begin{figure}[ht]
\centering
\includegraphics[width=\textwidth]
{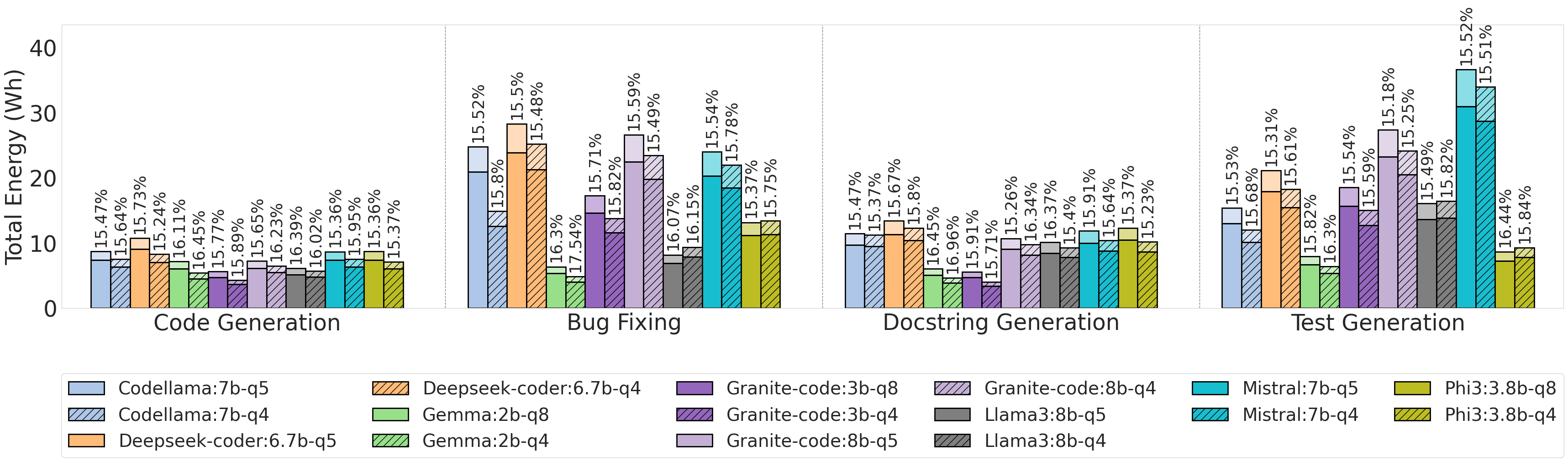}
\caption{Total energy (Wh) consumed by the RTX 3070 GPU and the Intel Core i7 CPU when performing each task. The lower part of each bar represents GPU energy, and the upper part represents CPU energy. Model names follow the pattern of Figure~\ref{fig:total_energy_a100}.}
\label{fig:total_energy_rtx3070}
\end{figure}

\textcolor{black}{We specifically focus on four representative software development tasks, namely \code, \bug, \doc, and \test, as they have been consistently identified as some of the most common AI-assisted programming activities in recent surveys~\cite{stackoverflow2025survey} and studies~\cite{watson2022systematic}. These tasks cover complementary types of development work: implementation, maintenance, documentation, and testing. Moreover, all four tasks are supported by the same HumanEvalPack problems, so it enables direct task-level energy comparisons using identical underlying problems~\cite{muennighoff2023octopack}.}
\\
To compare a model's energy consumption across different tasks, we report the GPU's total energy usage in watt-hours (Wh). Figures~\ref{fig:total_energy_a100} and~\ref{fig:total_energy_rtx3070} present the energy footprint of the analyzed models in the four tasks, for the two machines. In the remainder of the paper, except where explicitly noted, we focus on the A100 machine. Moreover, in this section, we are only concerned with energy consumption, as the use of these models incurs an energy cost independently of whether they are accurate or not. In Section~\ref{sec:f2} we also take model accuracy into account. 

In Figure~\ref{fig:total_energy_a100}, bars representing the same language model families, with the same number of parameters, have the same color. Full-precision (FP16) models are indicated by a striped pattern, Q8 models by plain bars, and Q4 models by a hatched pattern.
The figure shows that different tasks entail different energy use, and \test~and \bug~are the most energy-intensive tasks. The mean energy consumed across all the models for \code, \doc, \bug, and \test~was, respectively, 11.5Wh, 16.0Wh, 25.8Wh, and 32.9Wh. 
Column \textit{``Total energy (Wh)''} of Table~\ref{tab:a100_res_ee} presents numerical results for the energy consumption of the models across the tasks.\footnote{For the sake of readability, some non-significant models are omitted from Tables~\ref{tab:a100_res_ee},\ref{tab:a100_res_ac} and Pareto front plots in Fig~\ref{fig:a100_pareto}; however, they were included in the overall analysis. The corresponding numerical results can be found in Tables~\ref{tab:a100_res_app_ee},\ref{tab:a100_res_app_ac} in the Appendix (Section~\ref{sec:app}).} Numerical results for the RTX 3070 are available in the \textcolor{black}{online Appendix B~\cite{Alizadeh2026_replication}.}

The difference in the energy consumption of the same model for different tasks can be large. Model \textcolor{black}{Granite3.3:8b-fp16 consumed 4.9}$\times$ more energy for \test~than it did for \code, whereas Granite-code:20b-fp16 consumed 2.51$\times$ more energy for \bug~than it did for \doc. 
This is consistent for \test, although the difference is smaller in some cases. For example, \test~with Gemma:2b-fp16 consumed only 36.3\% more energy than \doc. There was no scenario where a model consumed more energy for \code~than \test. When considering \test~and \doc, the Phi3:3.8b family of models was the only one where the latter required more energy than the former (47.6\% more for Phi3:3.8b-q8).  
Comparing \bug~and \test, for $2/3$ of the model families, the latter required more energy than the former. The difference can be significant, e.g., Starcoder2:15b-q8 consumed 3$\times$ more energy when generating tests, compared to fixing bugs. Conversely, \bug~with Granite-code:20b-q8 required \textcolor{black}{74.61}\% more energy than \test. 
\bug~was generally more expensive than \code~and \doc. Only \textcolor{black}{Gemma:2b-q4} exhibited a slightly higher energy footprint for \code~than \bug. In addition, \textcolor{black}{three} families of models, Starcoder2:15b, Codellama:13b, and \textcolor{black}{Llama3:8b}, required consistently more energy for \doc~than \bug~and Phi3:14b, exhibited similar results for both tasks. Finally, \code~was the task requiring the least energy. \textcolor{black}{Five} model families consumed consistently more energy to generate code than to generate documentation: Codegemma:7b, Gemma:2b, Deepseek-coder:1.3b, \textcolor{black}{Gemma3n:e4b, and Qwen3:4b,} and two had very similar results, Gemma:7b and Granite-code:3b. 

Zooming into each task individually also reveals great variation in energy consumption. For example, Gemma:7b-fp16 consumed \textcolor{black}{8}$\times$ more energy than Granite-code:3b-q4 for \code. There is even more variation for the other tasks. The most energy-hungry model in \bug, Granite-code:20b-fp16, consumed almost \textcolor{black}{16}$\times$ more energy than the least energy-hungry, Gemma:2b-q4. In a similar vein, the most energy-hungry models for \doc~and \test, Starcoder2:15b-fp16 and Codegemma:7b-fp16, respectively, consumed \textcolor{black}{19.08}$\times$ and \textcolor{black}{15.18}$\times$ more energy than the least consuming ones in the respective tasks, \textcolor{black}{Granite3-moe:3b-q4} and Gemma:2b-q4. Overall, we have \textcolor{black}{six} different models occupying the extremes for the four tasks. If energy efficiency is an important quality attribute, considering the task at hand is essential.

Figure~\ref{fig:total_energy_a100} highlights two trends. \textbf{First}, for the same task, full-precision variants of a model family consistently used more energy than their quantized counterparts. Moreover, 8-bit quantized models exhibited higher energy consumption than their 4-bit counterparts. The few exceptions involved only quantized variants and the differences were relatively small, e.g., 2\% for \test~with Llama3:8b. Figure~\ref{fig:total_energy_rtx3070} shows a similar phenomenon. There are a few more exceptions (4 out of 32), but this is expected as the number of quantization bits is similar (5 vs. 4). \textbf{Second}, models with more parameters tend to consume more energy, but only for large differences, relative to model sizes, e.g., 15b vs. 7b models. When differences are smaller, the number of parameters is not a good predictor for the energy footprint of the models. For example, Phi3:14b-fp16 requires \textcolor{black}{8.10}\% more energy to generate docstrings than Granite-code:20b-fp16, but \textcolor{black}{59.28}\% less to fix bugs, and the two models are almost tied in \test.
\textcolor{black}{In contrast, for Mixture of Experts (MoE) models, the total number of parameters alone is no longer meaningful. Among the two MoE models considered, Deepseek-coder-v2:16b and Granite3-moe:3b, the former has 5.3x more parameters, yet its energy consumption for \code~and \bug~is only around 2x higher, and for \doc~it is less than 3x higher. On \test, their energy usage is essentially the same, and for the most aggressively quantized configuration (Q4), Deepseek-coder-v2 even consumes less energy.
The number of active parameters for Deepseek-coder-v2:16b and Granite3-moe:3b is approximately 2.4B and 800M, respectively. When compared to dense models with similar parameter counts, Deepseek-coder-v2:16b exhibits an even lower energy footprint on \code~than Gemma:2b and Deepseek-coder:1.3b. Although companies often highlight the total number of parameters to emphasize their models' capabilities, this can be misleading with respect to their actual energy footprint, as MoE models can have lower energy consumption comparable to that of much smaller dense models.
\\
Therefore, parameter count can only be a good estimator of energy consumption for dense models. When MoE models are included, one must instead consider the number of active parameters, as discussed further in Section~\ref{sec:f3}.
For instance, Gemma3n:e4b is a dense model that is listed with 4B effective parameters, which is lower than its total parameter count of 6.9B, as reported by Ollama. The \textit{``E''} prefix indicates that this model can operate with a reduced set of effective parameters. Google built a flexible parameter technology into Gemma3n that allows it to reduce compute and memory requirements during execution through parameter skipping and PLE caching techniques~\cite{gemma3n2025}. Accordingly, we consider 4B as its effective (active) parameters.
\\
Our results show that its memory usage is lower compared to models in the 6.7B–8B range (12 GB vs. 16 GB), yet higher than models with similar parameter counts, such as Phi3:3.8B and Qwen3:4B (12 GB vs. 10 GB and 9.7 GB, respectively). However, when we consider energy consumption, its usage is comparable to models in the 6.7B–8B range, and in some cases, it is even higher than models in the 13B–20B range. This underscores that relying on memory usage as a proxy for efficiency can be misleading, as it does not necessarily align with actual energy consumption.
}
\\
It is important to note that even different variants of the same model are not always comparable. For example, in Figure~\ref{fig:total_energy_a100}, although Codellama:13b-q8 and Codellama:7b-fp16 are close in size (14GB vs. 13GB), with the former having twice as many parameters at 8-bit precision and the latter having half as many parameters at full precision, their energy footprint differs. They show similar energy usage for code generation, but this does not hold for other tasks due to differences in their characteristics. This topic is explored further in Section~\ref{sec:f3}.
\\
For model inference, unlike model training, CPU energy may be significant, due to the need to constantly switch between CPU and GPU processing and the limited ability to batch multiple prompts. Since reading CPU energy via RAPL requires privileged access, and we did not have sudo access on the A100 machine, CPU energy data is only available for the RTX3070. As shown in Figure~\ref{fig:total_energy_rtx3070}, although all LLMs fit within the GPU’s memory capacity, i.e., no model processing had to be done on the CPU, CPU energy consumption still accounted for approximately 16\% of total energy usage, indicating a non-negligible contribution. This finding underscores the importance of considering CPU energy consumption, even when the primary workload is handled by the GPU, as it can contribute significantly to overall energy costs.
\\
\textcolor{black}{\textbf{Practical context.}
The energy values reported in Table~\ref{tab:a100_res_ee} are aggregated over all 164 benchmark problems. To estimate the average energy per query, these values should be divided by 164. For example, we consider Qwen3:4b-q8 because a 4B quantized model can easily fit on a local device. It consumes approximately 0.041, 0.075, 0.036, and 0.176~Wh per query for \code, \bug, \doc, and \test, respectively. In isolation, its average energy consumption across the four tasks (0.082~Wh per query) may seem negligible; however, its footprint can rapidly multiply in real world use.
\\
Modern development workflows increasingly rely on agentic coding assistants~\cite{anthropic2026agentic}, such as Claude Code and Cursor, which may call a model several times to complete one visible developer request. A recent study of around 4300 real-world Claude Code and Codex sessions found that an agent performs an average of 8.8 LLM calls to complete one user request~\cite{zhu2026tracelab}. Using this average and assuming that each internal call consumes roughly the same energy as an average query in our experiments, one \textit{agentic request} would consume approximately $0.082 \times 8.8 = 0.722$~Wh. For comparison, a recent 14-inch MacBook Pro has a battery capacity of 72.4~Wh~\cite{jary2026macbook}. Therefore, one full charge would be equivalent to the energy required for approximately 100 such \textit{agentic requests}, without accounting for screen, CPU, memory, and operating system overhead.
Although even the energy cost of one \textit{agentic request} is relatively small, it becomes non-negligible through repeated model calls and large-scale use.
\\
This comparison is only intended to put the measured energy into a more familiar context. It is not a prediction of actual MacBook battery life because our measurements were taken on NVIDIA GPUs and do not cover the full system.}

\begin{summary}
{\footnotesize 
\textbf{Summary.} 
Language models have varying energy footprints when performing diverse software development tasks. Even considering a code-specific model and two tasks that produce code, such as \code~and \bug, there may be a large difference in energy. Also, we have observed differences in energy usage of more than an order of magnitude between models performing the same task. The differences are not consistent across tasks. These trends could be observed for both machines. Finally, for inference, CPU energy can be non-negligible.

\vspace{3pt}
\noindent\textbf{Implications.} These findings highlight that the software development tasks that a model supports directly impact its energy use. Moreover, the differences in performance for different tasks can be large. Selecting models based on their expected tasks is crucial to reducing energy consumption. \textcolor{black}{In addition, MoE models break the simple link between total parameter count and energy usage, making the number of active parameters a more relevant indicator.}

}
\end{summary}

\begin{table*}[!htp]\centering
\begin{threeparttable}
\caption{GPU A100 - Comparative results of \textbf{\newtext{Energy and Efficiency}} in inference in Code Generation (CG), Bug Fixing (BF), Docstring Generation (DG) and Test Generation(TG)}\label{tab:a100_res_ee}
\scriptsize
\begin{tabular}{l@{\hskip 2pt}rrrr||rrrr}\toprule
\multirow{2}{*}{Model name} & \multicolumn{4}{c}{Total energy (Wh)} & \multicolumn{4}{c}{Efficiency (tokens/J)} \\
\cmidrule(lr){2-5}\cmidrule(lr){6-9}
 & CG & BF & DG & TG & CG & BF & DG & TG \\\midrule
\cellcolor[HTML]{E8E8E8}Codellama:13b-fp16\tnote{*}  &\cellcolor[HTML]{E8E8E8}18.91 &\cellcolor[HTML]{E8E8E8}25.56 &\cellcolor[HTML]{E8E8E8}35.52 &\cellcolor[HTML]{E8E8E8}53.84 &\cellcolor[HTML]{E8E8E8}0.20 &\cellcolor[HTML]{E8E8E8}0.19 &\cellcolor[HTML]{E8E8E8}0.20 &\cellcolor[HTML]{E8E8E8}0.20 \\
Codellama:13b-q8 &13.51 &19.03 &24.48 &41.24 &0.27 &0.26 &0.28 &0.28 \\
\cellcolor[HTML]{E8E8E8}Codellama:13b-q4 &\cellcolor[HTML]{E8E8E8}15.11 &\cellcolor[HTML]{E8E8E8}28.71 &\cellcolor[HTML]{E8E8E8}21.10 &\cellcolor[HTML]{E8E8E8}33.62 &\cellcolor[HTML]{E8E8E8}0.33 &\cellcolor[HTML]{E8E8E8}0.32 &\cellcolor[HTML]{E8E8E8}0.34 &\cellcolor[HTML]{E8E8E8}0.33 \\
Codellama:7b-fp16\tnote{*}  &11.15 &27.81 &17.52 &22.31 &0.36 &0.35 &0.36 &0.35 \\
\cellcolor[HTML]{E8E8E8}Codellama:7b-q8 &\cellcolor[HTML]{E8E8E8}8.17 &\cellcolor[HTML]{E8E8E8}20.23 &\cellcolor[HTML]{E8E8E8}12.43 &\cellcolor[HTML]{E8E8E8}15.91 &\cellcolor[HTML]{E8E8E8}0.50 &\cellcolor[HTML]{E8E8E8}0.50 &\cellcolor[HTML]{E8E8E8}0.52 &\cellcolor[HTML]{E8E8E8}0.50 \\
Codellama:7b-q4 &8.05 &17.16 &12.38 &14.09 &0.54 &0.54 &0.56 &0.53 \\
\cellcolor[HTML]{E8E8E8}Deepseek-coder-v2:16b-fp16\tnote{*}  &\cellcolor[HTML]{E8E8E8}7.18 &\cellcolor[HTML]{E8E8E8}24.67 &\cellcolor[HTML]{E8E8E8}13.70 &\cellcolor[HTML]{E8E8E8}17.73 &\cellcolor[HTML]{E8E8E8}0.57 &\cellcolor[HTML]{E8E8E8}0.57 &\cellcolor[HTML]{E8E8E8}0.57 &\cellcolor[HTML]{E8E8E8}0.55 \\
Deepseek-coder-v2:16b-q8 &5.72 &18.23 &10.26 &13.58 &0.73 &0.76 &0.76 &0.72 \\
\cellcolor[HTML]{E8E8E8}Deepseek-coder-v2:16b-q4 &\cellcolor[HTML]{E8E8E8}4.81 &\cellcolor[HTML]{E8E8E8}15.39 &\cellcolor[HTML]{E8E8E8}7.67 &\cellcolor[HTML]{E8E8E8}9.58 &\cellcolor[HTML]{E8E8E8}0.87 &\cellcolor[HTML]{E8E8E8}0.89 &\cellcolor[HTML]{E8E8E8}0.89 &\cellcolor[HTML]{E8E8E8}0.87 \\
Deepseek-coder:6.7b-fp16\tnote{*}  &15.64 &43.29 &20.50 &31.69 &0.36 &0.36 &0.36 &0.36 \\
\cellcolor[HTML]{E8E8E8}Deepseek-coder:6.7b-q8 &\cellcolor[HTML]{E8E8E8}10.72 &\cellcolor[HTML]{E8E8E8}30.26 &\cellcolor[HTML]{E8E8E8}14.31 &\cellcolor[HTML]{E8E8E8}22.30 &\cellcolor[HTML]{E8E8E8}0.51 &\cellcolor[HTML]{E8E8E8}0.50 &\cellcolor[HTML]{E8E8E8}0.52 &\cellcolor[HTML]{E8E8E8}0.50 \\
Deepseek-coder:6.7b-q4 &8.98 &29.37 &14.06 &22.11 &0.54 &0.54 &0.55 &0.54 \\
\cellcolor[HTML]{E8E8E8}Gemma3n:e4b-fp16 &\cellcolor[HTML]{E8E8E8}28.42 &\cellcolor[HTML]{E8E8E8}41.27 &\cellcolor[HTML]{E8E8E8}24.96 &\cellcolor[HTML]{E8E8E8}47.93 &\cellcolor[HTML]{E8E8E8}0.34 &\cellcolor[HTML]{E8E8E8}0.34 &\cellcolor[HTML]{E8E8E8}0.34 &\cellcolor[HTML]{E8E8E8}0.34 \\
Gemma3n:e4b-q8 &22.44 &32.43 &20.00 &37.48 &0.44 &0.43 &0.44 &0.43 \\
\cellcolor[HTML]{E8E8E8}Gemma3n:e4b-q4 &\cellcolor[HTML]{E8E8E8}20.01 &\cellcolor[HTML]{E8E8E8}28.65 &\cellcolor[HTML]{E8E8E8}15.49 &\cellcolor[HTML]{E8E8E8}27.14 &\cellcolor[HTML]{E8E8E8}0.49 &\cellcolor[HTML]{E8E8E8}0.49 &\cellcolor[HTML]{E8E8E8}0.50 &\cellcolor[HTML]{E8E8E8}0.49 \\
Gemma:2b-fp16 &8.90 &9.51 &7.96 &10.85 &0.79 &0.79 &0.78 &0.77 \\
\cellcolor[HTML]{E8E8E8}Gemma:2b-q8 &\cellcolor[HTML]{E8E8E8}7.39 &\cellcolor[HTML]{E8E8E8}8.19 &\cellcolor[HTML]{E8E8E8}6.34 &\cellcolor[HTML]{E8E8E8}8.76 &\cellcolor[HTML]{E8E8E8}0.97 &\cellcolor[HTML]{E8E8E8}0.94 &\cellcolor[HTML]{E8E8E8}0.99 &\cellcolor[HTML]{E8E8E8}0.97 \\
Gemma:2b-q4 &6.12 &\textbf{5.75} &5.21 &\textbf{7.58} &1.15 &1.12 &1.17 &1.14 \\
\cellcolor[HTML]{E8E8E8}Gemma:7b-fp16 &\cellcolor[HTML]{E8E8E8}34.29 &\cellcolor[HTML]{E8E8E8}53.23 &\cellcolor[HTML]{E8E8E8}34.29 &\cellcolor[HTML]{E8E8E8}58.55 &\cellcolor[HTML]{E8E8E8}0.28 &\cellcolor[HTML]{E8E8E8}0.27 &\cellcolor[HTML]{E8E8E8}0.28 &\cellcolor[HTML]{E8E8E8}0.28 \\
Gemma:7b-q8 &27.23 &42.10 &27.43 &47.98 &0.35 &0.34 &0.35 &0.34 \\
\cellcolor[HTML]{E8E8E8}Gemma:7b-q4 &\cellcolor[HTML]{E8E8E8}21.69 &\cellcolor[HTML]{E8E8E8}38.20 &\cellcolor[HTML]{E8E8E8}22.19 &\cellcolor[HTML]{E8E8E8}38.04 &\cellcolor[HTML]{E8E8E8}0.43 &\cellcolor[HTML]{E8E8E8}0.43 &\cellcolor[HTML]{E8E8E8}0.44 &\cellcolor[HTML]{E8E8E8}0.43 \\
Granite-code:20b-fp16\tnote{*}  &26.38 &95.47 &38.00 &54.35 &0.14 &0.13 &0.14 &0.13 \\
\cellcolor[HTML]{E8E8E8}Granite-code:20b-q8 &\cellcolor[HTML]{E8E8E8}18.92 &\cellcolor[HTML]{E8E8E8}67.54 &\cellcolor[HTML]{E8E8E8}26.89 &\cellcolor[HTML]{E8E8E8}38.68 &\cellcolor[HTML]{E8E8E8}0.19 &\cellcolor[HTML]{E8E8E8}0.18 &\cellcolor[HTML]{E8E8E8}0.19 &\cellcolor[HTML]{E8E8E8}0.18 \\
Granite-code:20b-q4 &15.78 &58.09 &21.05 &33.49 &0.22 &0.22 &0.23 &0.22 \\
\cellcolor[HTML]{E8E8E8}Granite-code:3b-fp16\tnote{*}  &\cellcolor[HTML]{E8E8E8}7.26 &\cellcolor[HTML]{E8E8E8}24.72 &\cellcolor[HTML]{E8E8E8}7.63 &\cellcolor[HTML]{E8E8E8}27.01 &\cellcolor[HTML]{E8E8E8}0.51 &\cellcolor[HTML]{E8E8E8}0.49 &\cellcolor[HTML]{E8E8E8}0.52 &\cellcolor[HTML]{E8E8E8}0.49 \\
Granite-code:3b-q8 &5.45 &18.05 &5.74 &19.62 &0.69 &0.67 &0.71 &0.67 \\
\cellcolor[HTML]{E8E8E8}Granite-code:3b-q4 &\cellcolor[HTML]{E8E8E8}4.28 &\cellcolor[HTML]{E8E8E8}14.97 &\cellcolor[HTML]{E8E8E8}3.92 &\cellcolor[HTML]{E8E8E8}16.42 &\cellcolor[HTML]{E8E8E8}0.85 &\cellcolor[HTML]{E8E8E8}0.82 &\cellcolor[HTML]{E8E8E8}0.88 &\cellcolor[HTML]{E8E8E8}0.83 \\
Granite-code:8b-fp16\tnote{*}  &10.89 &42.84 &16.75 &45.98 &0.29 &0.29 &0.29 &0.29 \\
\cellcolor[HTML]{E8E8E8}Granite-code:8b-q8 &\cellcolor[HTML]{E8E8E8}7.89 &\cellcolor[HTML]{E8E8E8}30.45 &\cellcolor[HTML]{E8E8E8}12.06 &\cellcolor[HTML]{E8E8E8}32.83 &\cellcolor[HTML]{E8E8E8}0.40 &\cellcolor[HTML]{E8E8E8}0.40 &\cellcolor[HTML]{E8E8E8}0.42 &\cellcolor[HTML]{E8E8E8}0.41 \\
Granite-code:8b-q4 &6.88 &27.82 &11.02 &28.75 &0.43 &0.43 &0.45 &0.44 \\
\cellcolor[HTML]{E8E8E8}Granite3-moe:3b-fp16 &\cellcolor[HTML]{E8E8E8}4.26 &\cellcolor[HTML]{E8E8E8}12.92 &\cellcolor[HTML]{E8E8E8}5.27 &\cellcolor[HTML]{E8E8E8}17.41 &\cellcolor[HTML]{E8E8E8}1.05 &\cellcolor[HTML]{E8E8E8}1.02 &\cellcolor[HTML]{E8E8E8}1.07 &\cellcolor[HTML]{E8E8E8}1.02 \\
Granite3-moe:3b-q8 &2.91 &8.62 &3.60 &11.24 &1.55 &1.51 &1.61 &1.50 \\
\cellcolor[HTML]{E8E8E8}Granite3-moe:3b-q4 &\cellcolor[HTML]{E8E8E8}\textbf{2.53} &\cellcolor[HTML]{E8E8E8}7.07 &\cellcolor[HTML]{E8E8E8}\textbf{2.68} &\cellcolor[HTML]{E8E8E8}10.04 &\cellcolor[HTML]{E8E8E8}1.80 &\cellcolor[HTML]{E8E8E8}1.84 &\cellcolor[HTML]{E8E8E8}1.92 &\cellcolor[HTML]{E8E8E8}1.84 \\
Granite3.3:2b-fp16 &6.47 &15.71 &10.46 &15.66 &0.66 &0.63 &0.66 &0.64 \\
\cellcolor[HTML]{E8E8E8}Granite3.3:2b-q8 &\cellcolor[HTML]{E8E8E8}4.43 &\cellcolor[HTML]{E8E8E8}10.89 &\cellcolor[HTML]{E8E8E8}7.07 &\cellcolor[HTML]{E8E8E8}10.17 &\cellcolor[HTML]{E8E8E8}0.94 &\cellcolor[HTML]{E8E8E8}0.92 &\cellcolor[HTML]{E8E8E8}0.96 &\cellcolor[HTML]{E8E8E8}0.92 \\
Granite3.3:2b-q4 &4.98 &10.23 &6.10 &10.85 &1.07 &1.06 &1.08 &1.08 \\
\cellcolor[HTML]{E8E8E8}Granite3.3:8b-fp16 &\cellcolor[HTML]{E8E8E8}12.18 &\cellcolor[HTML]{E8E8E8}47.47 &\cellcolor[HTML]{E8E8E8}22.18 &\cellcolor[HTML]{E8E8E8}59.59 &\cellcolor[HTML]{E8E8E8}0.29 &\cellcolor[HTML]{E8E8E8}0.29 &\cellcolor[HTML]{E8E8E8}0.29 &\cellcolor[HTML]{E8E8E8}0.29 \\
Granite3.3:8b-q8 &8.82 &34.47 &15.50 &43.02 &0.41 &0.41 &0.42 &0.41 \\
\cellcolor[HTML]{E8E8E8}Granite3.3:8b-q4 &\cellcolor[HTML]{E8E8E8}8.24 &\cellcolor[HTML]{E8E8E8}34.00 &\cellcolor[HTML]{E8E8E8}14.86 &\cellcolor[HTML]{E8E8E8}32.89 &\cellcolor[HTML]{E8E8E8}0.43 &\cellcolor[HTML]{E8E8E8}0.44 &\cellcolor[HTML]{E8E8E8}0.45 &\cellcolor[HTML]{E8E8E8}0.44 \\
Llama3.2:3b-fp16 &6.05 &11.88 &9.39 &28.68 &0.63 &0.61 &0.62 &0.62 \\
\cellcolor[HTML]{E8E8E8}Llama3.2:3b-q8 &\cellcolor[HTML]{E8E8E8}4.35 &\cellcolor[HTML]{E8E8E8}8.63 &\cellcolor[HTML]{E8E8E8}6.67 &\cellcolor[HTML]{E8E8E8}20.51 &\cellcolor[HTML]{E8E8E8}0.88 &\cellcolor[HTML]{E8E8E8}0.85 &\cellcolor[HTML]{E8E8E8}0.88 &\cellcolor[HTML]{E8E8E8}0.86 \\
Llama3.2:3b-q4 &3.59 &6.19 &5.75 &16.69 &1.05 &0.99 &1.05 &1.02 \\
\cellcolor[HTML]{E8E8E8}Llama3:8b-fp16 &\cellcolor[HTML]{E8E8E8}8.67 &\cellcolor[HTML]{E8E8E8}12.79 &\cellcolor[HTML]{E8E8E8}14.65 &\cellcolor[HTML]{E8E8E8}24.52 &\cellcolor[HTML]{E8E8E8}0.32 &\cellcolor[HTML]{E8E8E8}0.31 &\cellcolor[HTML]{E8E8E8}0.32 &\cellcolor[HTML]{E8E8E8}0.32 \\
Llama3:8b-q8 &6.33 &9.62 &10.00 &17.45 &0.43 &0.42 &0.45 &0.45 \\
\cellcolor[HTML]{E8E8E8}Llama3:8b-q4 &\cellcolor[HTML]{E8E8E8}6.29 &\cellcolor[HTML]{E8E8E8}9.33 &\cellcolor[HTML]{E8E8E8}10.26 &\cellcolor[HTML]{E8E8E8}19.59 &\cellcolor[HTML]{E8E8E8}0.47 &\cellcolor[HTML]{E8E8E8}0.45 &\cellcolor[HTML]{E8E8E8}0.49 &\cellcolor[HTML]{E8E8E8}0.48 \\
Phi3:14b-fp16\tnote{*}  &20.79 &38.87 &41.08 &51.50 &0.19 &0.19 &0.19 &0.19 \\
\cellcolor[HTML]{E8E8E8}Phi3:14b-q8 &\cellcolor[HTML]{E8E8E8}14.30 &\cellcolor[HTML]{E8E8E8}26.71 &\cellcolor[HTML]{E8E8E8}28.59 &\cellcolor[HTML]{E8E8E8}36.70 &\cellcolor[HTML]{E8E8E8}0.27 &\cellcolor[HTML]{E8E8E8}0.27 &\cellcolor[HTML]{E8E8E8}0.28 &\cellcolor[HTML]{E8E8E8}0.27 \\
Phi3:14b-q4 &12.62 &24.37 &22.32 &31.19 &0.32 &0.32 &0.33 &0.32 \\
\cellcolor[HTML]{E8E8E8}Phi3:3.8b-fp16\tnote{*}  &\cellcolor[HTML]{E8E8E8}11.15 &\cellcolor[HTML]{E8E8E8}22.28 &\cellcolor[HTML]{E8E8E8}18.51 &\cellcolor[HTML]{E8E8E8}13.24 &\cellcolor[HTML]{E8E8E8}0.53 &\cellcolor[HTML]{E8E8E8}0.52 &\cellcolor[HTML]{E8E8E8}0.53 &\cellcolor[HTML]{E8E8E8}0.52 \\
Phi3:3.8b-q8 &8.32 &16.71 &13.89 &9.41 &0.70 &0.68 &0.71 &0.69 \\
\cellcolor[HTML]{E8E8E8}Phi3:3.8b-q4 &\cellcolor[HTML]{E8E8E8}7.44 &\cellcolor[HTML]{E8E8E8}16.29 &\cellcolor[HTML]{E8E8E8}11.53 &\cellcolor[HTML]{E8E8E8}10.73 &\cellcolor[HTML]{E8E8E8}0.84 &\cellcolor[HTML]{E8E8E8}0.81 &\cellcolor[HTML]{E8E8E8}0.85 &\cellcolor[HTML]{E8E8E8}0.81 \\
Qwen2.5-coder:7b-fp16\tnote{*}  &9.81 &27.44 &18.06 &22.63 &0.31 &0.31 &0.32 &0.31 \\
\cellcolor[HTML]{E8E8E8}Qwen2.5-coder:7b-q8 &\cellcolor[HTML]{E8E8E8}7.52 &\cellcolor[HTML]{E8E8E8}20.50 &\cellcolor[HTML]{E8E8E8}13.32 &\cellcolor[HTML]{E8E8E8}16.86 &\cellcolor[HTML]{E8E8E8}0.41 &\cellcolor[HTML]{E8E8E8}0.41 &\cellcolor[HTML]{E8E8E8}0.43 &\cellcolor[HTML]{E8E8E8}0.41 \\
Qwen2.5-coder:7b-q4 &6.23 &17.99 &10.85 &14.52 &0.48 &0.49 &0.50 &0.49 \\
\cellcolor[HTML]{E8E8E8}Qwen3:4b-fp16 &\cellcolor[HTML]{E8E8E8}8.75 &\cellcolor[HTML]{E8E8E8}16.47 &\cellcolor[HTML]{E8E8E8}7.31 &\cellcolor[HTML]{E8E8E8}35.24 &\cellcolor[HTML]{E8E8E8}0.45 &\cellcolor[HTML]{E8E8E8}0.44 &\cellcolor[HTML]{E8E8E8}0.45 &\cellcolor[HTML]{E8E8E8}0.45 \\
Qwen3:4b-q8 &6.67 &12.21 &5.94 &28.88 &0.60 &0.57 &0.59 &0.60 \\
\cellcolor[HTML]{E8E8E8}Qwen3:4b-q4 &\cellcolor[HTML]{E8E8E8}5.58 &\cellcolor[HTML]{E8E8E8}11.11 &\cellcolor[HTML]{E8E8E8}4.96 &\cellcolor[HTML]{E8E8E8}17.09 &\cellcolor[HTML]{E8E8E8}0.68 &\cellcolor[HTML]{E8E8E8}0.68 &\cellcolor[HTML]{E8E8E8}0.71 &\cellcolor[HTML]{E8E8E8}0.70 \\
Starcoder2:15b-fp16\tnote{*}  &26.28 &27.31 &48.46 &86.10 &0.17 &0.16 &0.17 &0.17 \\
\cellcolor[HTML]{E8E8E8}Starcoder2:15b-q8 &\cellcolor[HTML]{E8E8E8}19.43 &\cellcolor[HTML]{E8E8E8}21.57 &\cellcolor[HTML]{E8E8E8}34.45 &\cellcolor[HTML]{E8E8E8}64.88 &\cellcolor[HTML]{E8E8E8}0.23 &\cellcolor[HTML]{E8E8E8}0.22 &\cellcolor[HTML]{E8E8E8}0.24 &\cellcolor[HTML]{E8E8E8}0.23 \\
Starcoder2:15b-q4 &16.15 &19.07 &29.54 &38.45 &0.28 &0.26 &0.29 &0.28 \\
\bottomrule
\end{tabular}
\begin{tablenotes}
    \footnotesize
    \item[*] Code-specific model families.
\end{tablenotes}
\end{threeparttable}
\end{table*}

\begin{table*}[!htp]\centering
\begin{threeparttable}
\caption{GPU A100 - Comparative results of \textbf{\newtext{Accuracy}} in inference in Code Generation (CG), Bug Fixing (BF), Docstring Generation (DG) and Test Generation(TG). SC and BC stand for statement and branch coverage of correct tests.}\label{tab:a100_res_ac}
\scriptsize
\begin{tabular}{lrrr|rrr}\toprule
\multirow{2}{*}{Model name} &\multicolumn{4}{c}{Accuracy\%}  & \multicolumn{1}{c}{SC\%} &\multicolumn{1}{c}{BC\%} \\\cmidrule(lr){2-5}\cmidrule(lr){6-7}
&CG &BF &DG &TG & & \\\midrule
\cellcolor[HTML]{E8E8E8}Codellama:13b-fp16\tnote{*}  &\cellcolor[HTML]{E8E8E8}37.19 &\cellcolor[HTML]{E8E8E8}18.90 &\cellcolor[HTML]{E8E8E8}45.12 &\cellcolor[HTML]{E8E8E8}68.90 &\cellcolor[HTML]{E8E8E8}90.50 &\cellcolor[HTML]{E8E8E8}77.52 \\
Codellama:13b-q8 &40.85 &18.90 &47.56 &66.46 &90.18 &76.64 \\
\cellcolor[HTML]{E8E8E8}Codellama:13b-q4 &\cellcolor[HTML]{E8E8E8}40.24 &\cellcolor[HTML]{E8E8E8}14.63 &\cellcolor[HTML]{E8E8E8}46.34 &\cellcolor[HTML]{E8E8E8}67.68 &\cellcolor[HTML]{E8E8E8}90.26 &\cellcolor[HTML]{E8E8E8}76.01 \\
Codellama:7b-fp16\tnote{*}  &41.46 &14.63 &46.95 &79.88 &83.77 &67.73 \\
\cellcolor[HTML]{E8E8E8}Codellama:7b-q8 &\cellcolor[HTML]{E8E8E8}40.85 &\cellcolor[HTML]{E8E8E8}15.85 &\cellcolor[HTML]{E8E8E8}48.78 &\cellcolor[HTML]{E8E8E8}79.88 &\cellcolor[HTML]{E8E8E8}82.71 &\cellcolor[HTML]{E8E8E8}66.49 \\
Codellama:7b-q4 &39.02 &17.68 &49.39 &\textbf{85.37} &83.57 &65.45 \\
\cellcolor[HTML]{E8E8E8}Deepseek-coder-v2:16b-fp16\tnote{*}  &\cellcolor[HTML]{E8E8E8}76.83 &\cellcolor[HTML]{E8E8E8}48.17 &\cellcolor[HTML]{E8E8E8}69.51 &\cellcolor[HTML]{E8E8E8}50.00 &\cellcolor[HTML]{E8E8E8}88.12 &\cellcolor[HTML]{E8E8E8}83.23 \\
Deepseek-coder-v2:16b-q8 &76.22 &47.56 &67.68 &45.73 &88.09 &82.75 \\
\cellcolor[HTML]{E8E8E8}Deepseek-coder-v2:16b-q4 &\cellcolor[HTML]{E8E8E8}70.73 &\cellcolor[HTML]{E8E8E8}\textbf{50.61} &\cellcolor[HTML]{E8E8E8}63.41 &\cellcolor[HTML]{E8E8E8}46.34 &\cellcolor[HTML]{E8E8E8}89.64 &\cellcolor[HTML]{E8E8E8}85.10 \\
Deepseek-coder:6.7b-fp16\tnote{*}  &55.49 &31.71 &62.80 &62.80 &71.79 &57.34 \\
\cellcolor[HTML]{E8E8E8}Deepseek-coder:6.7b-q8 &\cellcolor[HTML]{E8E8E8}54.26 &\cellcolor[HTML]{E8E8E8}31.10 &\cellcolor[HTML]{E8E8E8}63.41 &\cellcolor[HTML]{E8E8E8}60.37 &\cellcolor[HTML]{E8E8E8}73.58 &\cellcolor[HTML]{E8E8E8}62.02 \\
Deepseek-coder:6.7b-q4 &54.26 &35.37 &67.68 &56.71 &79.03 &67.43 \\
\cellcolor[HTML]{E8E8E8}Gemma3n:e4b-fp16 &\cellcolor[HTML]{E8E8E8}56.10 &\cellcolor[HTML]{E8E8E8}25.00 &\cellcolor[HTML]{E8E8E8}73.78 &\cellcolor[HTML]{E8E8E8}27.44 &\cellcolor[HTML]{E8E8E8}72.61 &\cellcolor[HTML]{E8E8E8}81.78 \\
Gemma3n:e4b-q8 &58.54 &25.61 &\textbf{76.21} &26.22 &70.85 &81.82 \\
\cellcolor[HTML]{E8E8E8}Gemma3n:e4b-q4 &\cellcolor[HTML]{E8E8E8}55.49 &\cellcolor[HTML]{E8E8E8}23.17 &\cellcolor[HTML]{E8E8E8}69.51 &\cellcolor[HTML]{E8E8E8}26.83 &\cellcolor[HTML]{E8E8E8}73.57 &\cellcolor[HTML]{E8E8E8}79.71 \\
Gemma:2b-fp16 &18.29 &17.07 &68.90 &68.29 &82.81 &65.29 \\
\cellcolor[HTML]{E8E8E8}Gemma:2b-q8 &\cellcolor[HTML]{E8E8E8}18.29 &\cellcolor[HTML]{E8E8E8}15.85 &\cellcolor[HTML]{E8E8E8}67.07 &\cellcolor[HTML]{E8E8E8}70.12 &\cellcolor[HTML]{E8E8E8}82.86 &\cellcolor[HTML]{E8E8E8}64.53 \\
Gemma:2b-q4 &17.68 &16.46 &70.73 &73.17 &79.19 &57.52 \\
\cellcolor[HTML]{E8E8E8}Gemma:7b-fp16 &\cellcolor[HTML]{E8E8E8}25.00 &\cellcolor[HTML]{E8E8E8}8.53 &\cellcolor[HTML]{E8E8E8}67.07 &\cellcolor[HTML]{E8E8E8}22.56 &\cellcolor[HTML]{E8E8E8}75.56 &\cellcolor[HTML]{E8E8E8}75.79 \\
Gemma:7b-q8 &26.83 &6.70 &68.90 &24.39 &75.13 &75.19 \\
\cellcolor[HTML]{E8E8E8}Gemma:7b-q4 &\cellcolor[HTML]{E8E8E8}26.22 &\cellcolor[HTML]{E8E8E8}8.53 &\cellcolor[HTML]{E8E8E8}75.61 &\cellcolor[HTML]{E8E8E8}22.56 &\cellcolor[HTML]{E8E8E8}77.26 &\cellcolor[HTML]{E8E8E8}80.15 \\
Granite-code:20b-fp16\tnote{*}  &62.20 &30.49 &50.61 &60.98 &92.00 &78.92 \\
\cellcolor[HTML]{E8E8E8}Granite-code:20b-q8 &\cellcolor[HTML]{E8E8E8}60.98 &\cellcolor[HTML]{E8E8E8}31.71 &\cellcolor[HTML]{E8E8E8}48.78 &\cellcolor[HTML]{E8E8E8}60.37 &\cellcolor[HTML]{E8E8E8}92.96 &\cellcolor[HTML]{E8E8E8}80.30 \\
Granite-code:20b-q4 &59.76 &31.71 &46.34 &60.98 &91.05 &78.85 \\
\cellcolor[HTML]{E8E8E8}Granite-code:3b-fp16\tnote{*}  &\cellcolor[HTML]{E8E8E8}45.73 &\cellcolor[HTML]{E8E8E8}20.73 &\cellcolor[HTML]{E8E8E8}49.39 &\cellcolor[HTML]{E8E8E8}76.22 &\cellcolor[HTML]{E8E8E8}69.47 &\cellcolor[HTML]{E8E8E8}39.12 \\
Granite-code:3b-q8 &45.12 &20.12 &51.22 &77.44 &68.82 &38.52 \\
\cellcolor[HTML]{E8E8E8}Granite-code:3b-q4 &\cellcolor[HTML]{E8E8E8}43.90 &\cellcolor[HTML]{E8E8E8}15.24 &\cellcolor[HTML]{E8E8E8}47.56 &\cellcolor[HTML]{E8E8E8}73.17 &\cellcolor[HTML]{E8E8E8}72.80 &\cellcolor[HTML]{E8E8E8}45.74 \\
Granite-code:8b-fp16\tnote{*}  &43.90 &20.12 &57.93 &60.98 &70.86 &42.58 \\
\cellcolor[HTML]{E8E8E8}Granite-code:8b-q8 &\cellcolor[HTML]{E8E8E8}43.90 &\cellcolor[HTML]{E8E8E8}20.73 &\cellcolor[HTML]{E8E8E8}58.54 &\cellcolor[HTML]{E8E8E8}62.80 &\cellcolor[HTML]{E8E8E8}70.82 &\cellcolor[HTML]{E8E8E8}42.37 \\
Granite-code:8b-q4 &37.19 &18.29 &60.37 &66.46 &70.38 &41.49 \\
\cellcolor[HTML]{E8E8E8}Granite3-moe:3b-fp16 &\cellcolor[HTML]{E8E8E8}29.88 &\cellcolor[HTML]{E8E8E8}9.14 &\cellcolor[HTML]{E8E8E8}50.00 &\cellcolor[HTML]{E8E8E8}29.26 &\cellcolor[HTML]{E8E8E8}72.01 &\cellcolor[HTML]{E8E8E8}76.88 \\
Granite3-moe:3b-q8 &32.32 &7.31 &48.78 &31.09 &72.38 &74.92 \\
\cellcolor[HTML]{E8E8E8}Granite3-moe:3b-q4 &\cellcolor[HTML]{E8E8E8}28.05 &\cellcolor[HTML]{E8E8E8}9.14 &\cellcolor[HTML]{E8E8E8}43.90 &\cellcolor[HTML]{E8E8E8}48.78 &\cellcolor[HTML]{E8E8E8}77.26 &\cellcolor[HTML]{E8E8E8}70.60 \\
Granite3.3:2b-fp16 &53.66 &32.32 &54.87 &73.17 &49.35 &26.42 \\
\cellcolor[HTML]{E8E8E8}Granite3.3:2b-q8 &\cellcolor[HTML]{E8E8E8}53.66 &\cellcolor[HTML]{E8E8E8}32.32 &\cellcolor[HTML]{E8E8E8}56.10 &\cellcolor[HTML]{E8E8E8}73.17 &\cellcolor[HTML]{E8E8E8}51.93 &\cellcolor[HTML]{E8E8E8}30.08 \\
Granite3.3:2b-q4 &46.95 &27.44 &59.15 &68.90 &49.96 &32.81 \\
\cellcolor[HTML]{E8E8E8}Granite3.3:8b-fp16 &\cellcolor[HTML]{E8E8E8}59.76 &\cellcolor[HTML]{E8E8E8}33.54 &\cellcolor[HTML]{E8E8E8}65.85 &\cellcolor[HTML]{E8E8E8}11.58 &\cellcolor[HTML]{E8E8E8}71.65 &\cellcolor[HTML]{E8E8E8}77.09 \\
Granite3.3:8b-q8 &60.98 &31.10 &68.29 &14.63 &70.36 &73.36 \\
\cellcolor[HTML]{E8E8E8}Granite3.3:8b-q4 &\cellcolor[HTML]{E8E8E8}64.02 &\cellcolor[HTML]{E8E8E8}28.05 &\cellcolor[HTML]{E8E8E8}71.95 &\cellcolor[HTML]{E8E8E8}18.90 &\cellcolor[HTML]{E8E8E8}74.86 &\cellcolor[HTML]{E8E8E8}78.02 \\
Llama3.2:3b-fp16 &50.61 &27.44 &56.10 &10.37 &66.66 &63.65 \\
\cellcolor[HTML]{E8E8E8}Llama3.2:3b-q8 &\cellcolor[HTML]{E8E8E8}51.83 &\cellcolor[HTML]{E8E8E8}29.27 &\cellcolor[HTML]{E8E8E8}53.65 &\cellcolor[HTML]{E8E8E8}10.37 &\cellcolor[HTML]{E8E8E8}64.90 &\cellcolor[HTML]{E8E8E8}62.92 \\
Llama3.2:3b-q4 &50.61 &27.44 &50.61 &19.51 &63.03 &60.16 \\
\cellcolor[HTML]{E8E8E8}Llama3:8b-fp16 &\cellcolor[HTML]{E8E8E8}54.26 &\cellcolor[HTML]{E8E8E8}40.85 &\cellcolor[HTML]{E8E8E8}50.00 &\cellcolor[HTML]{E8E8E8}31.71 &\cellcolor[HTML]{E8E8E8}83.66 &\cellcolor[HTML]{E8E8E8}80.35 \\
Llama3:8b-q8 &57.32 &40.85 &51.83 &34.15 &85.41 &82.08 \\
\cellcolor[HTML]{E8E8E8}Llama3:8b-q4 &\cellcolor[HTML]{E8E8E8}52.44 &\cellcolor[HTML]{E8E8E8}39.02 &\cellcolor[HTML]{E8E8E8}48.78 &\cellcolor[HTML]{E8E8E8}27.44 &\cellcolor[HTML]{E8E8E8}79.80 &\cellcolor[HTML]{E8E8E8}79.39 \\
Phi3:14b-fp16\tnote{*}  &62.80 &40.24 &68.29 &52.44 &80.35 &68.28 \\
\cellcolor[HTML]{E8E8E8}Phi3:14b-q8 &\cellcolor[HTML]{E8E8E8}62.19 &\cellcolor[HTML]{E8E8E8}39.02 &\cellcolor[HTML]{E8E8E8}70.73 &\cellcolor[HTML]{E8E8E8}49.39 &\cellcolor[HTML]{E8E8E8}78.96 &\cellcolor[HTML]{E8E8E8}66.90 \\
Phi3:14b-q4 &48.78 &34.15 &68.90 &45.73 &70.71 &55.11 \\
\cellcolor[HTML]{E8E8E8}Phi3:3.8b-fp16\tnote{*}  &\cellcolor[HTML]{E8E8E8}59.15 &\cellcolor[HTML]{E8E8E8}23.78 &\cellcolor[HTML]{E8E8E8}56.71 &\cellcolor[HTML]{E8E8E8}65.85 &\cellcolor[HTML]{E8E8E8}75.46 &\cellcolor[HTML]{E8E8E8}55.95 \\
Phi3:3.8b-q8 &56.10 &20.73 &55.49 &60.98 &76.07 &55.65 \\
\cellcolor[HTML]{E8E8E8}Phi3:3.8b-q4 &\cellcolor[HTML]{E8E8E8}48.78 &\cellcolor[HTML]{E8E8E8}14.63 &\cellcolor[HTML]{E8E8E8}57.32 &\cellcolor[HTML]{E8E8E8}49.39 &\cellcolor[HTML]{E8E8E8}78.64 &\cellcolor[HTML]{E8E8E8}57.77 \\
Qwen2.5-coder:7b-fp16\tnote{*}  &75.00 &46.95 &75.00 &54.26 &91.99 &83.75 \\
\cellcolor[HTML]{E8E8E8}Qwen2.5-coder:7b-q8 &\cellcolor[HTML]{E8E8E8}76.22 &\cellcolor[HTML]{E8E8E8}45.73 &\cellcolor[HTML]{E8E8E8}75.00 &\cellcolor[HTML]{E8E8E8}51.22 &\cellcolor[HTML]{E8E8E8}91.73 &\cellcolor[HTML]{E8E8E8}83.57 \\
Qwen2.5-coder:7b-q4 &61.58 &39.63 &75.00 &57.32 &92.17 &83.90 \\
\cellcolor[HTML]{E8E8E8}Qwen3:4b-fp16 &\cellcolor[HTML]{E8E8E8}\textbf{84.75} &\cellcolor[HTML]{E8E8E8}45.12 &\cellcolor[HTML]{E8E8E8}74.39 &\cellcolor[HTML]{E8E8E8}29.88 &\cellcolor[HTML]{E8E8E8}73.73 &\cellcolor[HTML]{E8E8E8}84.98 \\
Qwen3:4b-q8 &\textbf{84.75} &45.12 &73.78 &34.15 &79.36 &83.20 \\
\cellcolor[HTML]{E8E8E8}Qwen3:4b-q4 &\cellcolor[HTML]{E8E8E8}82.31 &\cellcolor[HTML]{E8E8E8}32.32 &\cellcolor[HTML]{E8E8E8}75.00 &\cellcolor[HTML]{E8E8E8}29.88 &\cellcolor[HTML]{E8E8E8}74.30 &\cellcolor[HTML]{E8E8E8}85.18 \\
Starcoder2:15b-fp16\tnote{*}  &61.58 &42.07 &68.90 &75.00 &71.10 &47.59 \\
\cellcolor[HTML]{E8E8E8}Starcoder2:15b-q8 &\cellcolor[HTML]{E8E8E8}59.76 &\cellcolor[HTML]{E8E8E8}41.46 &\cellcolor[HTML]{E8E8E8}70.73 &\cellcolor[HTML]{E8E8E8}65.85 &\cellcolor[HTML]{E8E8E8}69.69 &\cellcolor[HTML]{E8E8E8}46.01 \\
Starcoder2:15b-q4 &58.54 &40.85 &71.95 &79.88 &64.87 &37.61 \\
\bottomrule
\end{tabular}
\begin{tablenotes}
    \footnotesize
    \item[*] Code-specific model families.
\end{tablenotes}
\end{threeparttable}
\end{table*}
\subsection{\textbf{When it comes to energy consumption and accuracy of LLMs in software development, it is not always a compromise. }}\label{sec:f2}

Although energy consumption is an important quality attribute, the utility of a model is determined by its accuracy when performing inference. \textcolor{black}{This is particularly important in software engineering tasks, where incorrect outputs are often not only suboptimal but also unusable, as they may fail to compile, fail tests, or introduce unsafe behavior. As a result, energy-accuracy trade-offs have practical implications when selecting models for local development workflows.}

In this section, we investigate the relationship between the accuracy of the analyzed models in the 4 tasks and their energy footprint. Since the ideal scenario is one with low energy usage and high accuracy, we employed a Pareto front (or ``Pareto frontier'') analysis to determine the trade-off between these two objectives. \textcolor{black}{The Pareto frontier method determines the best solution sets for multiple performance indices~\cite{yang2022review}.}
In our case, it identifies models that offer the best balance between minimizing energy usage and maximizing accuracy. By visualizing our results using this approach, we aim to highlight optimal solutions where improvements in one objective do not come at the expense of the other.

\begin{figure*}[htb]
    \centering
    \begin{minipage}[b]{0.49\textwidth}
        \centering
        \includegraphics[width=\textwidth]{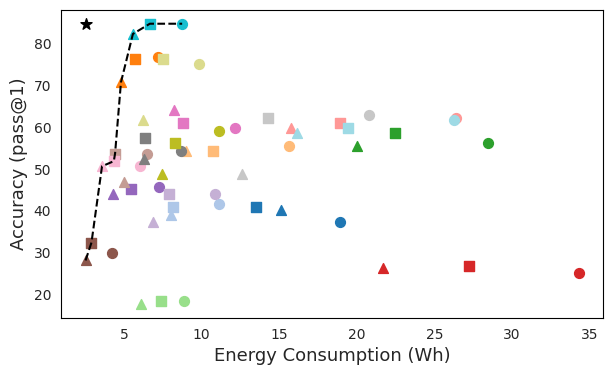}
    \subcaption*{(a) Code Generation}
    \end{minipage}
    \hfill
    \begin{minipage}[b]{0.49\textwidth}
        \centering
        \includegraphics[width=\textwidth]{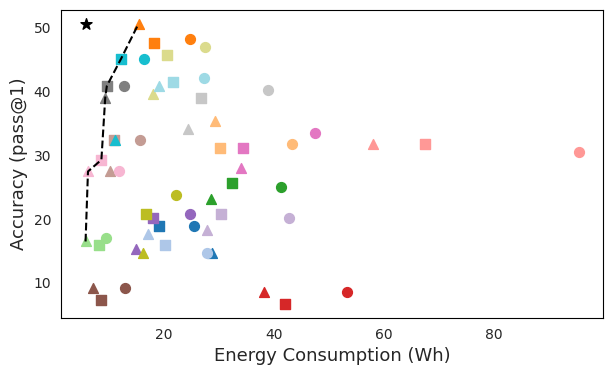}
    \subcaption*{(b) Bug Fixing}
    \end{minipage}
    
    \vspace{0.1cm}
    
    \begin{minipage}[b]{0.49\textwidth}
        \centering
        \includegraphics[width=\textwidth]{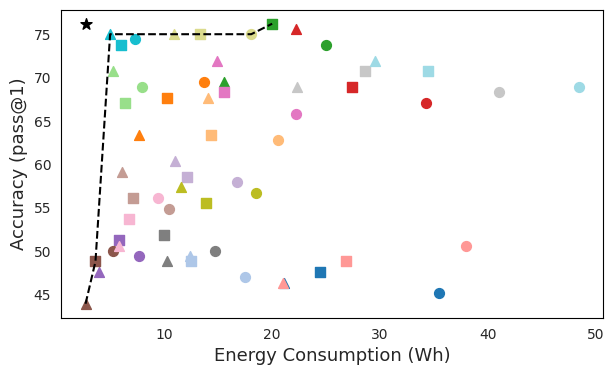}
    \subcaption*{(c) Docstring Generation}
    \end{minipage}
    \hfill
    \begin{minipage}[b]{0.49\textwidth}
        \centering
        \includegraphics[width=\textwidth]{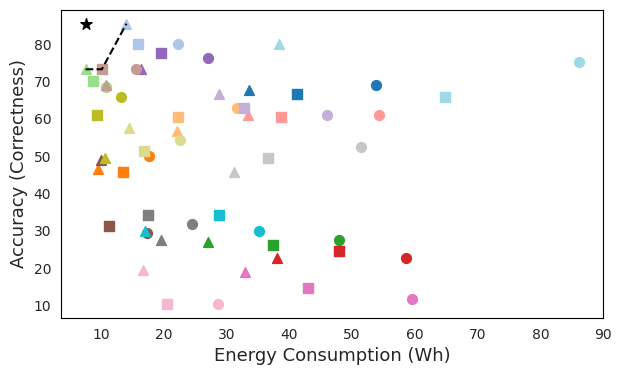}
    \subcaption*{(d) Test Generation}
    \end{minipage}
    \begin{minipage}[b]{\textwidth}
        \centering
        \includegraphics[width=\textwidth]{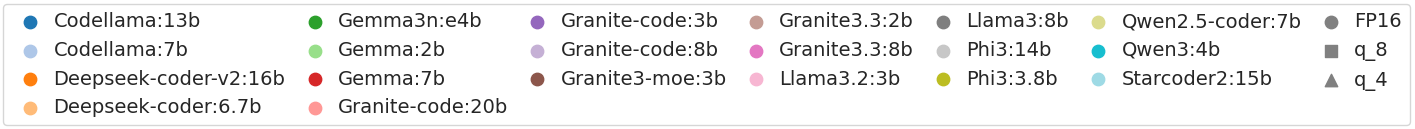}
    \end{minipage}
    
    \caption{Relation between accuracy and total energy for each task on GPU A100. The black dashed line indicates the Pareto frontier. The \textbf{\ding{72}} marker denotes the ideal point. (Data points exceeding the y-axis limit are not displayed.)}
\label{fig:a100_pareto}
\end{figure*}

Figure~\ref{fig:a100_pareto}(a) presents the results for \code. The best results are the ones closer to the top left of the plot. It represents the overall lowest energy consumption and highest accuracy across all the models for this specific task. The Pareto front is indicated by the black dashed line. Models in that line are not dominated, i.e., there is no other model that has at the same time higher accuracy and lower energy consumption. 
\textcolor{black}{The models on the Pareto frontier are Granite3-moe:3b-q4, Granite3-moe:3b-q8, Llama3.2:3b-q4, Llama3.2:3b-q8, Granite3.3:2b-q8, Deepseek-coder-v2:16b-q4, Qwen3:4b-q4, Qwen3:4b-q8, Qwen3:4b-fp16.}
The figure shows that \textcolor{black}{almost all models on the Pareto frontier are general-purpose models, with the only exception being Deepseek-coder-v2, a coding-specialized MoE model. All dense code-specialized models were dominated by others in terms of the energy–accuracy trade-off. Furthermore, the models on the frontier are mostly quantized, with one exception, Qwen3:4b-fp16, whose accuracy matches that of its 8-bit quantized counterpart.}
\\
\textcolor{black}{
Another notable observation is that Qwen3:4b consistently outperforms all other models, including larger ones, in terms of accuracy, while maintaining a lower energy footprint. Large models such as Phi3:14b, Starcoder2:15b, and Granite-code:20b in their full-precision versions consume 2.37x, 3.00x, and 3.01x more energy, respectively, yet offer no improvement in accuracy.
\\
The model with the lowest accuracy on the Pareto frontier is Granite3-moe:3b-q4, the other MoE model in our collection. Despite having only 800M active parameters, it outperforms larger dense models such as Gemma:2b and Gemma:7b in terms of accuracy by 10.37\% and 1.83\%, respectively. 
}
\\
\textcolor{black}{
Similarly, in \bug~(Figure~\ref{fig:a100_pareto}(b)), all models on the Pareto frontier are quantized. \textcolor{black}{The Pareto set consists of Gemma:2b-q4, Llama3.2:3b-q4, Llama3.2:3b-q8, Llama3:8b-q4, Llama3:8b-q8, Qwen3:4b-q8, Deepseek-coder-v2:16b-q4.} Once again, general-purpose models dominate overall, with the single exception being the 4-bit quantized version of Deepseek-coder-v2, which achieves the highest accuracy with only a small increase in energy usage compared to Qwen3:4b-q8.
}

\textcolor{black}{
In contrast, in \doc~(Figure~\ref{fig:a100_pareto}(c)) all three variants of Qwen2.5-coder:7b appear on the Pareto frontier, demonstrating performance competitive with general-purpose models. \textcolor{black}{The list of all models on the Pareto frontier is Granite3-moe:3b-q4, Granite3-moe:3b-q8, Qwen3:4b-q4, Qwen2.5-coder:7b-q4, Qwen2.5-coder:7b-q8, Qwen2.5-coder:7b-fp16, Gemma3n:e4b-q8} 
\\
The steep slope at the beginning of the Pareto fronts in Figures~\ref{fig:a100_pareto}(a–c) indicates that accuracy can be significantly improved with a relatively small increase in energy consumption, emphasizing the trade-offs involved in model selection. Figure~\ref{fig:a100_pareto}(c) further shows that small increases in energy consumption yield large accuracy gains at first, but eventually reach a point of diminishing returns, where additional energy brings only marginal improvements.
\\
This diminishing return in accuracy relative to energy usage emphasizes that an increase in model size does not always correlate with proportional gains in energy efficiency. Models such as Gemma3n:4b-q8 and Qwen2.5:7b family also illustrate this trend, requiring considerable energy for slight accuracy gains.
\\
A noteworthy case for \doc~is Qwen3:4b. The three models of this family are positioned near the ideal point. In particular, Qwen3:4b-q4, a small model that could be run locally on a smartphone, is only surpassed in accuracy (75\%) by Gemma3n:e4b-q8 (76.21\%), while the latter model consumed 2.74x more energy.
}

The Pareto frontier for \test~(Figure~\ref{fig:a100_pareto}(d)) is relatively short \textcolor{black}{and consists of Gemma:2b-q4, Granite3.3:2b-q8, Codellama:7b-q4}. This shows that few solutions are close to the ideal point in terms of both energy consumption and accuracy for this task. There are only \textcolor{black}{three} models in the frontier: Gemma:2b-q4, Granite3.3:2b-q8, and Codellama:7b-q4. The latter consumes 1.86x and 1.39x more energy than Gemma:2b-q4 and Granite3.3:2b-q8, respectively, and has a 12.2\% higher accuracy. Overall, Codellama:7b-q4 exhibits the best accuracy across all the models in this task \textcolor{black}{while maintaining relatively low energy consumption.}

There is also variation in the energy footprint of the same model when performing different tasks. \textcolor{black}{For instance, Qwen2.5-coder:7b-fp16 is on the Pareto frontier for \doc, has reasonable performance for \code~and \bug, and performs poorly for \test.}

Significant variation also exists when examining how multiple models perform the same task. \textcolor{black}{For example, one of the most energy-hungry models when generating tests on the A100, Starcoder2:15b-fp16, consumed 11.35x more energy than the least energy-hungry, Gemma:2b-q4. Differences in energy footprint are not directly connected to differences in accuracy. The accuracy of Gemma:2b-q4 for \test~was 73.17\% vs. 75\% for Starcoder2:15b-fp16, even though the former is significantly smaller in the number of parameters and uses aggressive quantization.}

\textcolor{black}{Even variants of the same model family can exhibit noticeable differences in performing the same task. The model Granite-code:20b-fp16, when generating docstring, consumed 4.95x more energy than its smaller, quantized version, Granite-code:3b-fp16.}
Complementarily, the two aforementioned variants of Granite-code exhibited comparable accuracy when producing documentation, 50.61\%, and 49.39\%, respectively.

\textcolor{black}{Examining each model (data points with the same color) across different quantization levels (indicated by different markers) shows that moving quantized (Q4 and Q8) to full precision (FP16) offers only marginal or sometimes even negative accuracy gains at a substantial energy cost. In the plots, it is easy to see this by identifying the large number of squares and triangles (representing quantized models) that appear higher along the Y-axis, closer to the ideal point. Across all four tasks, quantized models took the top spot in accuracy, with the exception of \code, where two models, one 8-bit quantized and one FP16, achieved the same highest accuracy.}

Regardless of accuracy, our findings indicate that the smallest models with the most severe quantization level (Q4) consistently exhibited the lowest inference times across all four tasks on both hardware setups, with \textbf{\code} being the fastest task. More details for both GPUs can be found in online appendices C and D~\cite{Alizadeh2026_replication}.

These results highlight the importance of choosing the right model for the right task if it is going to be used intensively. This diversity of options allows practitioners to select models that align with specific requirements, optimizing for either energy efficiency, accuracy, or a combination of both. For example, in contexts where high accuracy is essential but energy costs are a concern, \textcolor{black}{Qwen3:4b} could be prioritized. Conversely, if energy is less of a constraint and marginal accuracy improvements are critical, larger models could be considered.

\textcolor{black}{Finally, coding-specific models outperform general-purpose models in \test~and \bug, with no general model appearing among the top five in terms of accuracy. For \code, if we set aside Qwen3:4b (a general model) and its configurations, which clearly cut above the rest, the next top five models are all coding-specific. In \doc, however, Gemma:7b-q4 and Gemma3n:e4b-q8 rank in the top two positions, in a close competition with coding-specific models. This underscores the importance of task-aware model selection. Because \doc~is closer to natural language, general models can compete more effectively with coding-specific models on this task.}

\begin{summary}
{\footnotesize 
\textbf{Summary.} 
Larger models often have a much greater energy footprint while usually performing close to or even being outperformed by smaller models in terms of accuracy. We observed that small increases in energy consumption result in substantial accuracy improvements, but then a point of diminishing returns is reached, emphasizing that an increase in model size does not always correlate with proportional gains in accuracy. Additionally, quantized models sometimes exhibit higher accuracy than their full-precision counterparts while having much lower model sizes.

\vspace{3pt}
\noindent\textbf{Implications.} The results show that not all models follow a strict trade-off pattern between accuracy and energy use when considering different software development tasks. In some cases, a model that consumes (relatively) little energy also exhibits a high accuracy, and it is not necessary to prioritize one attribute to the detriment of the other, i.e., it is not necessary to find a compromise solution.

}
\end{summary}

\subsection{\textbf{Characteristics of a model's architecture have a strong correlation with its efficiency measured in the number of output tokens per joule.}}\label{sec:f3}

\begin{figure}[tb]
\centerline{
\includegraphics[width=0.8\linewidth]
{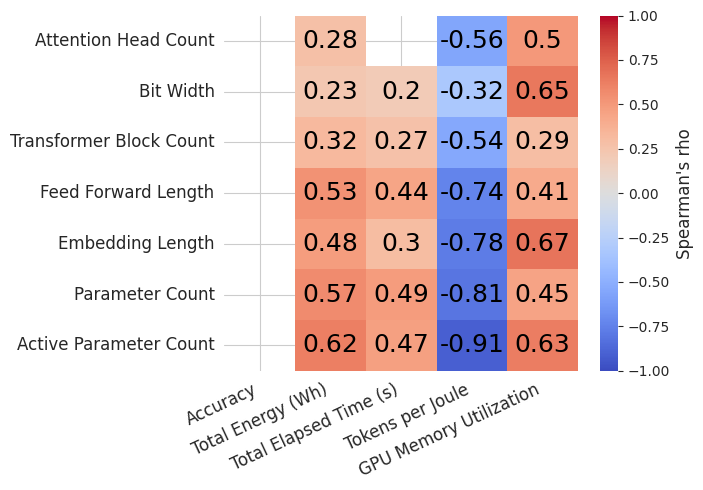}}
\caption{Spearman's correlation matrix for all models across all tasks on GPU A100 ($p-value < 0.00143$)}
\label{fig:correlation_matrix}
\end{figure}

Figure~\ref {fig:correlation_matrix} shows a correlation matrix between \textcolor{black}{seven} characteristics of a model's architecture and five performance metrics. The numbers in the cells indicate the Spearman's correlation coefficient between two variables, e.g., \textcolor{black}{-0.32} is the correlation between bit width and (output) tokens/J. \textcolor{black}{We used Spearman's correlation as it does not assume linear relationships or normality.} The matrix only presents coefficients for which the p-value is statistically significant, e.g., the correlation between Parameter Count and Accuracy is not statistically significant. All the values in the correlation matrix in Figure~\ref{fig:correlation_matrix} are based on an adjusted alpha value. To control the family-wise error rate, we applied the Bonferroni correction by dividing the default alpha value (0.05) by the number of conducted tests \textcolor{black}{(35)}. All the numbers in the matrix have a p-value less than \textcolor{black}{0.00143}.

This figure shows that parameter count \textcolor{black}{or even active parameter count} does not have a statistically significant correlation with model accuracy. This result underscores the findings of Section~\ref{sec:f2}, that larger models do not necessarily produce more accurate results than smaller ones. \textcolor{black}{Although the parameter count shows a strong negative correlation with the tokens/J efficiency metric, the correlation with active parameter count is even stronger. This indicates that, for MoE models, only the fraction of parameters activated during inference meaningfully contributes to energy usage.}
For tokens/J, a high value is better, as it quantifies the efficiency of a model, instead of just its resource usage. 
\textcolor{black}{We report energy efficiency as tokens per joule (tokens/J), representing the amount of output produced per unit of energy; this is the reciprocal of joules per token (J/token), which reflects the energy cost per generated token. As a result, correlations with tokens/J have the opposite direction compared to J/token, while maintaining the same magnitude and conveying the same underlying relationship. In the remainder of this paper, we use tokens/J as our primary measure of efficiency for consistency.}
\\
\textcolor{black}{In this context, the strong negative correlation indicates that models with a larger number of active parameters are less efficient (in tokens/J), corresponding to a higher energy cost per output token (J/token).}
\textcolor{black}{Additionally, the metric tokens/J is negatively correlated with every analyzed model characteristic. Although this correlation is weak for bit width.} The ``Efficiency (tokens/J)'' column of Table~\ref{tab:a100_res_ee} also indicates that tokens/J may be a useful metric to analyze model performance across different software development tasks. Despite the observed variation in overall energy consumption across different tasks, the table shows a small variation in the value of the metric, per model, regardless of task.
Active parameter count also has a moderate (total time) to strong (total energy and GPU memory) positive correlation with the remaining performance metrics. The remaining model characteristics exhibit statistically significant correlations with the performance metrics (except for accuracy), ranging from weak to strong. \textcolor{black}{except for between the number of attention heads with total elapsed time.
Finally, there is no statistically significant correlation between architectural attributes and accuracy. These findings confirm that larger models tend to be slower, consume more energy, and require more memory, but this does not guarantee higher accuracy.
}

The strong positive correlation between memory usage and bit width confirms that lower-precision models require less memory bandwidth and storage, reducing the energy required for data transfer and memory access.
Results in Table~\ref{tab:a100_res_ee} indicate that FP16 variants have lower efficiency in tokens/J than 8-bit or 4-bit variants. The 4-bit variant of each model exhibited the highest performance in tokens/J for all tasks and the lowest energy usage in \textcolor{black}{307} out of the \textcolor{black}{312} cases analyzed (26 model families, 3 variants, and 4 tasks).

As shown in the Figure.~\ref{fig:correlation_matrix}, Energy consumption is positively correlated with key hyperparameters, such as the number of attention heads, feed-forward network dimension, number of transformer blocks, embedding size, and overall model size. These hyperparameters vary across models, influencing their energy usage. We also observe that energy consumption is moderately correlated with the token count in the generated output \textcolor{black}{($statistic = 0.65, \; p < 10^{-37}$)}. \textcolor{black}{In contrast, the correlation with input tokens is weaker ($statistic = 0.43, \; p < 10^{-14}$), suggesting that output token generation has a stronger impact on energy consumption than input processing.}
In all the analyzed cases, inference time and overall energy usage have an almost perfect correlation \textcolor{black}{($statistic = 0.96, \; p < 10^{-171}$)}.

\textcolor{black}{
To relate these observations to theoretical time complexity, inspired by~\cite{mcintosh2019can}, we examine proxy terms that reflect the expected scaling behavior of autoregressive transformer inference. As depicted in Figure~\ref{fig:correlation_matrix}, tokens per joule are strongly correlated with the number of active parameters, indicating that the energy cost per generated token increases with model size. This observation is consistent with the expected behavior of transformer inference, where each generated token requires a forward pass through all active parameters. Consequently, the overall inference cost is expected to scale with both the number of generated tokens and the number of active parameters. Based on this intuition, we compute Spearman's correlations between energy consumption and complexity proxies defined as the product of parameter count with output tokens, input tokens, and total tokens. The results show a strong correlation ($statistic = 0.92$, \; $p < 10^{-130}$) for the output-based term ($output\_tokens \times active\_parameter\_count$), compared to input-based ($statistic = 0.71$, \; $p < 10^{-47}$) and total-token-based ($statistic = 0.83$, \; $p < 10^{-80}$) formulations.
\\
These findings suggest that energy consumption is most closely aligned with a complexity proxy of the form $O(np)$, where $n$ denotes the number of generated tokens and $p$ the number of active parameters. However, as these aggregate terms do not distinguish the individual contributions of input and output tokens, in Section~\ref{sec:f5} we complement this analysis with regression models to separately estimate their effects.
}

\begin{summary}
{\footnotesize 
\textbf{Summary.} 
\textcolor{black}{The performance metrics, except for accuracy, are correlated with the architectural characteristics of the models. Furthermore, among these metrics, tokens/J exhibits the highest correlation coefficients, particularly with the active parameter count. Our results also show that energy consumption is most strongly associated with the joint effect of output length and active parameter count, supporting a complexity proxy of the form $O(np)$. In addition, the consistent tokens/J observed across all four tasks highlight it as an effective metric for evaluating the efficiency of the models.}

\vspace{3pt}
\noindent\textbf{Implications.} LLM architecture, together with the expected size of the output produced for a given task, can help us estimate a model's energy efficiency when performing software development tasks. \textcolor{black}{In particular, the number of generated tokens and the number of active parameters jointly provide a useful indicator of energy consumption.} Model size is not a reliable predictor of a model's accuracy in software development tasks.

}
\end{summary}


\subsection{\textbf{Coding models are not particularly advantageous for coding tasks, if energy consumption matters.}}\label{sec:f4}

According to our results, there is no relationship between a model's energy consumption and whether it is code-specific or general-purpose. Our analysis reveals that when we only consider accuracy, coding-specific LLMs on average outperform general-purpose LLMs in 3 out of the 4 tasks evaluated. Coding models achieve higher accuracy compared to general ones for \code, \bug, and \test, with mean accuracies of 50.57\% vs. 37.62\%, 27.52\% vs. 21.73\%, and 56.83\% vs. 34.49\%, respectively. For \doc, the performance is comparable, with mean accuracies of 57.93\% and 57.89\%, respectively.

A careful examination of the results paints a more nuanced picture. For \bug, the accuracy of the models was generally low; the highest value reached only 50.61\%. When we factor in energy consumption, most models in the Pareto frontier (Figure~\ref{fig:a100_pareto}(b)) are non-coding models (Gemma:2b and Llama3:8b, Llama3.2:3b and Qwen3:4b), the exception being Deepseek-coder-v2:16b-q4, the model with the highest accuracy in this task.

Similarly, for \code, Deepseek-coder-v2:16b-q4 is the only coding model on the Pareto frontier (Figure~\ref{fig:a100_pareto}(a)).
In \doc~(Figure~\ref{fig:a100_pareto}(c)), all configurations of Qwen2.5-coder:7b appear on the Pareto frontier, and they all achieve identical accuracy. The accuracy of Qwen2.5-coder:7b-q4 is only 1.21\% lower than the best, while consuming 54.25\% of the best model’s energy.

Moreover, for \test~(Figure~\ref{fig:a100_pareto}(d)), there are only three models in the Pareto frontier, two of them, Gemma:2b-q4 and Granite3.3:2b-q8, are non-coding models. Furthermore, there is a small cluster of models near the Pareto frontier, and the others consume much more energy, have much lower accuracy, or both.

\textcolor{black}{
To statistically analyze the results, we apply the Mann-Whitney U test to compare coding-specific and general-purpose models within each task. For each task, we conduct two tests: one for accuracy and one for energy consumption. This results in a total of eight tests, therefore, we applied the Bonferroni Correction and adjusted the significance threshold $\alpha$ to 0.00625. The null hypothesis here ($H_0$) assumes that the distributions of the two groups are identical and it is rejected when a statistically significant difference is observed.
\\
The results show that coding models only achieve significantly higher accuracy in \test~(median 60.98\% vs. 29.88\%, $p = 0.00001$). For the other tasks, no statistically significant differences are observed in either accuracy or energy consumption.\\
This may raise the concern that the observed improvement in accuracy is driven by confounding factors such as larger model size or token usage. To address this, we employ a linear regression model with interaction terms between model type and task, while controlling for model size, input and output token counts, and quantization to isolate the effect of model type, and we select \doc~as the reference category. Following our findings in Section~\ref{sec:f3} for energy consumption, we further extend the model by including an interaction term between output tokens and model size to reflect the expected $O(np)$ computational complexity.
\\
The regression results indicate that neither model type nor its interaction with task has a statistically significant effect on energy consumption (see Appendix, Section~\ref{sec:app}, Table~\ref{tab:reg_energy_coding_general}). This indicates that the energy usage of coding models does not systematically vary from general models.
\\
In contrast, coding models significantly outperform general models in \code~(+13.14, $p = 0.028$), \bug~(+9.03, $p = 0.047$), and \test~(+18.78, $p = 0.02$), while no significant difference is observed in \doc~($p = 0.760$) (see Appendix, Section~\ref{sec:app}, Table~\ref{tab:reg_acc_coding_general}).
These findings confirm that general models can perform comparably to coding models in tasks involving natural language generation, whereas coding-specific models provide accuracy improvements in tasks that require a deeper understanding of code. However, these improvements are not accompanied by gains in energy efficiency.
}

\begin{summary}
{\footnotesize 
\textbf{Summary.}
\textcolor{black}{
After controlling for model size, token usage, and quantization, regression analysis shows that coding-specific models achieve significantly higher accuracy than general-purpose models in \code, \bug, and \test. In contrast, no statistically significant differences in energy consumption are observed between coding and general models across tasks. 
Additionally, for \code, \doc, and \test, some general-purpose models achieve both high accuracy and low energy consumption, while no model fared particularly well in \bug.}

\vspace{3pt}
\noindent\textbf{Implications.} \textcolor{black}{
Our findings indicate that coding-specific models provide genuine accuracy benefits in tasks requiring code understanding, even after accounting for confounding factors such as model size and token usage. However, these improvements do not translate into better energy efficiency as they rarely appear on the Pareto frontier when
energy is taken into account. This highlights the need for developing coding models that not only improve accuracy but also optimize computational efficiency across tasks.}

}
\end{summary}


\section{\textbf{In-depth Analysis}}\label{sec:f5}

In this section, we perform a more in-depth investigation from a systems perspective to help us better make sense of the results of Section~\ref{sec:res}. We start out by building a regression model to establish a relationship between the size and quantization levels of the models, the input (as determined by the respective tokenizers), and output lengths for each task, and the energy that is consumed as a result. We then proceed to investigate how intensively different CUDA kernel calls are used and what the impact of different tasks and quantization levels is.
\vspace{0.5cm}

\subsection{\textbf{Examining energy usage in terms of model and task characteristics}}
\label{sec:regression}

In this section, we examine the relationship between input and output token counts, model characteristics, and the total energy consumption for each task.
Because different models employ different tokenizers, the same input prompt can result in different token counts. Therefore, we report the overall average number of tokens per prompt, computed across all 78 models and divided by 164 prompts for each task. The average number of input tokens is 172.75 for \doc, 230.95 for \code, 317.93 for \test, and 377.13 for \bug. These variations reflect the different amounts of information required for each task. Although \bug~has the largest average input size, \test~shows the highest mean energy consumption across models (Section~\ref{sec:f1}). Similarly, \doc~has a smaller average input size than \code, yet its mean energy consumption is higher.
The average number of output tokens per prompt is 110.04, 144.26, 239.93, and 295.49 for \code, \doc, \bug, and \test, respectively. Notably, the order of output length aligns with the observed energy consumption patterns across tasks (Section~\ref{sec:f1}).
\textcolor{black}{These results are consistent with the findings in Section~\ref{sec:res} and with the complexity proxy of $O(np)$. However, there are cases where the relationship does not hold strictly. For instance, in~\test, although gemma3n:e4b-fp16 has 22M fewer active parameters than qwen3:4b-fp16 and generates slightly fewer tokens (56505 vs. 56849), it consumes 36\% more energy. This discrepancy may be attributed to its 27\% larger input size (57727 vs. 45529).
}

Therefore, to further investigate the relative contributions of input tokens and output tokens to total energy, we fit a separate multiple linear regression model in each task.
Across all tasks, the number of \textit{total output tokens} has a statistically larger spread than \textit{total input tokens}. The ratio of variances is 10.29$\times$ for \code, 11.40$\times$ for \bug, 26.98$\times$ in \test, and 6.11$\times$ in \doc.
Since the variables are on different scales, comparisons of raw (unstandardized) coefficients can be difficult and misleading. Therefore, we standardized input and output token counts, as well as the active parameter count (in billion parameters), using z-scores before fitting the model. This standardization allows the regression coefficients for input tokens, output tokens, and parameter count to be interpreted on a directly comparable scale.

We have also included the active parameter count in the model as a proxy for the model architecture, since it was identified as the most influential factor on energy consumption (Figure~\ref{fig:correlation_matrix}). Including parameter count allows the model to capture the relationship between token counts and energy usage more accurately. Other architectural characteristics were excluded because they are highly correlated with parameter count, and their inclusion could introduce multicollinearity, thereby violating linear regression assumptions.

Finally, we also encoded quantization level numerically by its bit width (2, 1, and 0.5 for FP16, Q8, and Q4, respectively) and incorporated it as a predictor in our regression model, as models with lower bit widths generally consume less energy. 
Since all quantization variants (FP16, Q8, and Q4) within a single family share the identical architectural characteristics, we grouped them under a common \textit{family label}, making 26 distinct model families.
To account for the residual correlation within each family, we employed Cluster\-Robust Standard Error~(CSE), using that \textit{family label} as the clustering variable. This approach provides consistent statistical inference and ensures p-values remain robust to shared variance and noise among related models.

The Linearity assumption was reasonably met, as the residuals were centered around zero with no apparent non-linear pattern in the Residuals vs. Fitted Values plots in the Appendix (Section~\ref{sec:app}, Figure~\ref{fig:linearity_diagnosis}). The Normality assumption was also reasonably met, \textcolor{black}{as the residuals appear approximately normally distributed, with minor deviations in the tails} as shown by the Q-Q plot and histogram of residuals in the Appendix (Section~\ref{sec:app}, Figure~\ref{fig:linearity_diagnosis}).
The general regression model is defined as shown in Equation~\ref{eq:regression_model}.

\begin{equation}
E = \beta_0 + \beta_1 I + \beta_2 O + \beta_3 P + \beta_4 I \cdot P + \beta_5 O \cdot P + \beta_6 Q + \varepsilon
\label{eq:regression_model}
\end{equation}
\\
\noindent where $E$ stands for the total energy consumption (Wh), $I$ is the standardized input token count, $O$ is the standardized output token count, $P$ represents the standardized active parameter count (in billions of parameters), $Q$ denotes the quantization level (bit width), \textcolor{black}{$IP$ and $OP$ represent interaction terms, and $\varepsilon$ is the error term.
Interaction terms are included to capture that the marginal energy cost of processing tokens depends on model size. In particular, the interaction between output tokens and active parameter count reflects the expected complexity term $O(np)$.}

The estimated coefficients, standard errors, and significance levels for each task are reported in Table~\ref{tab:regression_results}.
Because the four regressions share overlapping model characteristics (parameter count and quantization), we adjusted p-values using the Bonferroni method to account for potential dependencies across tasks. 
The default $\alpha$ value (0.05) is divided by the number of regressions, resulting in a corrected $\alpha$ of 0.0125.

\begin{table*}[tb]
\centering
\caption{Regression coefficients ($\beta$), standard errors (SE), statistical significance ($p-value < 0.0125$), and model fit metrics for each task. Cluster-Robust SEs are used (clustered by family). I$_{std}$, O$_{std}$, and P$_{std}$ denote standardized input tokens, output tokens, and parameter count, respectively.}
\label{tab:regression_results}
\renewcommand{\arraystretch}{0.95}
\begin{tabular}{lrrrrr}
\toprule
\multicolumn{6}{c}{\textbf{Code Generation}} \\
\midrule
\textbf{Predictor} & \textbf{Coefficient ($\beta$)} & \textbf{SE} & \textbf{$p$-value} & \textbf{$R^2$} & \textbf{Adj. $R^2$} \\
\midrule
Intercept              & 8.58  & 0.60  & $<$0.001 & \multirow{7}{*}{0.901} & \multirow{7}{*}{0.893} \\
Input Tokens$_{std}$   & 0.22  & 0.41  & 0.590    &  &  \\
Output Tokens$_{std}$  & 6.55  & 0.40  & $<$0.001 &  &  \\
Parameter Count$_{std}$& 6.23  & 0.27  & $<$0.001 &  &  \\
$I_{std} P_{std}$      & -0.35 & 0.40  & 0.385    &  &  \\
$O_{std} P_{std}$      & 3.89  & 0.50  & $<$0.001 &  &  \\
Quantization Level     & 3.54  & 0.42  & $<$0.001 &  &  \\
\midrule
\multicolumn{6}{c}{\textbf{Bug Fixing}} \\
\midrule
\textbf{Predictor} & \textbf{Coefficient ($\beta$)} & \textbf{SE} & \textbf{$p$-value} & \textbf{$R^2$} & \textbf{Adj. $R^2$} \\
\midrule
Intercept              & 17.95 & 1.29  & $<$0.001 & \multirow{7}{*}{0.925} & \multirow{7}{*}{0.919} \\
Input Tokens$_{std}$   & -0.20 & 0.53  & 0.710    &  &  \\
Output Tokens$_{std}$  & 9.91  & 0.51  & $<$0.001 &  &  \\
Parameter Count$_{std}$& 13.38 & 0.53  & $<$0.001 &  &  \\
$I_{std} P_{std}$      & -0.10 & 0.65  & 0.880    &  &  \\
$O_{std} P_{std}$      & 5.66  & 0.51  & $<$0.001 &  &  \\
Quantization Level     & 8.06  & 1.01  & $<$0.001 &  &  \\
\midrule
\multicolumn{6}{c}{\textbf{Docstring Generation}} \\
\midrule
\textbf{Predictor} & \textbf{Coefficient ($\beta$)} & \textbf{SE} & \textbf{$p$-value} & \textbf{$R^2$} & \textbf{Adj. $R^2$} \\
\midrule
Intercept              & 9.75  & 0.82  & $<$0.001 & \multirow{7}{*}{0.924} & \multirow{7}{*}{0.918} \\
Input Tokens$_{std}$   & 0.24  & 0.59  & 0.682    &  &  \\
Output Tokens$_{std}$  & 3.66  & 0.33  & $<$0.001 &  &  \\
Parameter Count$_{std}$& 7.54  & 0.34  & $<$0.001 &  &  \\
$I_{std} P_{std}$      & -0.33 & 0.62  & 0.594    &  &  \\
$O_{std} P_{std}$      & 1.78  & 0.21  & $<$0.001 &  &  \\
Quantization Level     & 5.25  & 0.67  & $<$0.001 &  &  \\
\midrule
\multicolumn{6}{c}{\textbf{Test Generation}} \\
\midrule
\textbf{Predictor} & \textbf{Coefficient ($\beta$)} & \textbf{SE} & \textbf{$p$-value} & \textbf{$R^2$} & \textbf{Adj. $R^2$} \\
\midrule
Intercept              & 21.38 & 1.62  & $<$0.001 & \multirow{7}{*}{0.942} & \multirow{7}{*}{0.937} \\
Input Tokens$_{std}$   & -0.26 & 0.86  & 0.758    &  &  \\
Output Tokens$_{std}$  & 14.11 & 0.52  & $<$0.001 &  &  \\
Parameter Count$_{std}$& 16.52 & 0.79  & $<$0.001 &  &  \\
$I_{std} P_{std}$      & -0.88 & 0.89  & 0.319    &  &  \\
$O_{std} P_{std}$      & 7.63  & 0.64  & $<$0.001 &  &  \\
Quantization Level     & 10.33 & 1.31  & $<$0.001 &  &  \\
\bottomrule
\end{tabular}
\end{table*}

\textcolor{black}{ 
As shown in Table~\ref{tab:regression_results}, \textbf{the regression model has high explanatory power.} \code~has the lowest adjusted $R^2$, with the model explaining 89.3\% of the variance in the energy. For \test, the model can explain 93.7\% of the variance, with \doc~and \bug~sitting in between, with 91.8\% and 91.9\%.
\\
Active parameter count, number of output tokens, and quantization level are the predictors whose coefficients are statistically significant across tasks. In contrast, the coefficient for the number of input tokens is both small in magnitude and not statistically significant, indicating that input length has a negligible impact on total energy consumption in our dataset, after controlling for other factors.
In addition, the interaction between output tokens and parameter count is statistically significant across all tasks, confirming that the effect of output length increases with model size. This finding aligns with the expected scaling behavior of autoregressive inference, where generating each token requires a forward pass through the model.
\\
Among the predictors, active parameter count and output tokens exhibit the strongest effects, except for \doc, where quantization ranks second after parameter count. 
For \test, the coefficient for output tokens is notably larger than other tasks, indicating that energy consumption is more strongly driven by output length. This can be attributed both to the nature of the task, as producing comprehensive and nontrivial unit tests involves generating long, structured outputs~\cite{schafer2023empirical, yang2024evaluation}, and to the high variation in output length (Table~\ref{tab:std_mean}), which amplifies the influence of output tokens on energy consumption.
In contrast, for \doc, the low variation in output length reduces the influence of output tokens, making quantization comparatively more influential factor.}

\textcolor{black}{Interpretation of the standardized coefficients requires accounting for interaction effects. For example, the coefficient $\beta_2 = 6.55$ for the \code~task represents the effect of output tokens when the parameter count is at its mean ($P_{std} = 0$). Due to the interaction term, the marginal effect of output tokens increases with model size. Given that the standard deviation of output tokens in this task is 8812.99 tokens, an increase of approximately 8813 output tokens corresponds to an increase of 6.55 Wh at the average model size. For a model with one standard deviation larger in parameter count, this effect increases to approximately 10.44~Wh ($6.55+3.89$), while for a smaller model it decreases accordingly.
Given that both the number of tokens and the total energy consumption are aggregated across 164 prompts, this effect can be interpreted at the prompt level as follows: at the average model size, an increase of approximately 53.7 ($8813/164$) output tokens per prompt is associated with an additional 144~J ($6.55\times3600/164$) of energy consumption per prompt.
\\
Similarly, in the \bug~task, where $\beta_3 = 13.38$, an increase of one standard deviation in the number of active parameters (4695.90M parameters) corresponds to an increase of approximately 13.38~Wh in total energy consumption, holding all other predictors constant at their mean values. Given that the reported values are aggregated over 164 prompts, this translates to an additional 294~J of energy per prompt.}
\\
Mean, median and standard deviations of token counts per task, computed across 78 models, are shown in the original scale to support the interpretation of standardized coefficients in Table~\ref{tab:std_mean}.
\begin{table}[ht]
\centering
\caption{Mean, Median and Standard Deviation of token-based predictors (aggregated over 164 prompts). The active parameter count is constant across tasks (Mean = 7210.20M parameters, Median = 6825.44 and SD = 4695.90M).}
\label{tab:std_mean}
\renewcommand{\arraystretch}{0.95}
\begin{tabular}{lrrrr}
\toprule
\textbf{Task} & \textbf{Variable} & \textbf{Mean} & \textbf{Median} & \textbf{SD} \\
\midrule
\multirow{2}{*}{Code Generation}
 & Input Tokens  & 37{,}875.10 & 37{,}827.00 & 2{,}747.71 \\
 & Output Tokens & 18{,}045.87 & 14{,}893.50 & 8{,}812.99 \\
\midrule
\multirow{2}{*}{Bug Fixing}
 & Input Tokens  & 61{,}849.96 & 61{,}793.00 & 4{,}322.62 \\
 & Output Tokens & 39{,}348.21 & 42{,}355.00 & 14{,}596.77 \\
\midrule
\multirow{2}{*}{Test Generation}
 & Input Tokens  & 52{,}140.35 & 52{,}055.50 & 3{,}815.53 \\
 & Output Tokens & 48{,}461.15 & 44{,}975.00 & 19{,}820.20 \\
\midrule
\multirow{2}{*}{Docstring Generation}
 & Input Tokens  & 28{,}331.72 & 28{,}439.50 & 2{,}220.16 \\
 & Output Tokens & 23{,}659.35 & 23{,}765.00 & 5{,}486.19 \\
\bottomrule
\end{tabular}
\end{table}

\textcolor{black}{\textbf{Implications.} Although developers cannot change these coefficients, the results can help them identify the main sources of energy use and make more informed decisions when selecting models. In particular, in our experiments, energy consumption is mainly driven by active parameter count, output length, and quantization, while the effect of output length becomes stronger for larger models. Therefore, for output-intensive tasks such as \test, developers can reduce energy use by selecting a model with fewer active parameters, applying stronger quantization, or limiting unnecessarily long outputs, given that the required accuracy is maintained. The differences between tasks also suggest opportunities for task-specific optimization instead of using the same model configuration for every workflow. Finally, the weak effect of input length should be interpreted in the context of its limited variation in our dataset.}

\subsection{\textbf{Usage of CUDA kernel calls}}\label{sec:cuda}

To better understand how quantization and different tasks influence workload distribution, we examined kernel-level CUDA activities during inference. This granular analysis reveals which specific GPU operations (e.g., matrix multiplications, memory transfers) dominate resource usage under different conditions.
For this purpose, we used NVIDIA Nsight Systems (nsys)~\cite{nvidiansight:2025} in command-line mode to profile CUDA activities. Specifically, we launched the Ollama server within the nsys profiling command, allowing us to capture the entire lifecycle of each run, from model loading, through the full inference process, to unloading. The exact command used for profiling is available in the online Appendix \cite{Alizadeh2026_replication}. Our analysis here focuses on time, instead of energy, since it is possible to monitor activity at a finer level of granularity. Also, the two variables exhibited almost perfect correlation in our study (Section~\ref{sec:f3}).

We filtered the models under examination to select those with an accuracy higher than the median. Across the four tasks, only three families appeared in this list for all of their variants: Deepseek-coder:6.7B, Starcoder2:15B, and Qwen2.5-coder:7B.

Before discussing kernel-level behavior, it is important to note that during inference, LLMs employ key–value (KV) caching so that each token’s representation is computed once and reused at later steps, avoiding repeated processing of earlier tokens. Table~\ref{tab:nsys_deepseek} illustrates the kernel distribution for Deepseek-coder:6.7B. For example, in the \code~task, the full-precision version, the top five most frequently invoked kernels correspond to matrix-vector multiplication, RMS normalization, writing in KV cache, and element-wise addition. The matrix-vector multiplication kernel is dominant, accounting for first, 81.7\% and then 2.9\% more of the total runtime, while RMS normalization, KV cache write, and element-wise addition contribute 3.6\%, 1.9\%, and 1.7\%, respectively. Tables for the remaining two LLMs are presented in the Appendix Section (\ref{sec:app}) in Tables~\ref{tab:nsys_qwen}, \ref{tab:nsys_starcoder}.
\\
For quantized models, matrix-vector multiplication remains the most frequent operation, but its relative share of execution time decreases; 66.2\% in the Q8 model and 64\% in the Q4 model, since part of the time is now spent on quantization. In the case of Q4, which applies mixed quantization (with some weights quantized to 6 or 5 bits and the rest to 4 bits), we observe two distinct matrix-vector multiplication kernels corresponding to different quantization levels (type 12 and type 14). More aggressive quantization requires slightly more processing time, with quantization accounting for 5.9\% of the total runtime in Q4 and 5.6\% in Q8. After the quantization steps, RMS normalization becomes the next most time-consuming operation.
While quantization effectively reduces memory traffic, it is not entirely free, and its quantization kernels appear among the top-3 contributors.
\\
If we look across different tasks, the matrix-vector multiplication kernel (\texttt{mul\_mat\_vec}) is still consistently on top, accounting for more than 64\% of the total runtime. This reflects the autoregressive nature of LLM inference, where every new token is generated through repeated multiplication of vectors in weight matrices. The type of task only affects the secondary kernels; for example, during \bug, the KV cache write becomes the third most time-consuming operation, while in the rest of the tasks, matrix multiplications dominate. This shows that task characteristics (longer inputs or longer outputs) can only shift the secondary kernels in the GPU workload mix.
\\
\begin{table*}[!htp]\centering
\begin{threeparttable}
\caption{GPU A100 – Distribution of kernel execution time (\%) across quantization levels (FP16, Q8, Q4) for different software engineering tasks when running \texttt{DeepSeek-coder:6.7B}.}
\label{tab:nsys_deepseek}
\scriptsize
\begin{tabular}{l r | l r | l r}
\toprule
\multicolumn{2}{c|}{FP16} & \multicolumn{2}{c|}{Q8} & \multicolumn{2}{c}{Q4} \\
Kernel & (\%) & Kernel & (\%) & Kernel & (\%) \\
\midrule

\multicolumn{6}{c}{Code Generation} \\
\cmidrule(r){1-6}
\texttt{mul\_mat\_vec\_f\_256} & 81.7 & \texttt{mul\_mat\_vec\_q\_8} & 66.2 & \texttt{mul\_mat\_vec\_q\_12} & 49.1 \\
\texttt{rms\_norm\_f32} & 3.6 & \texttt{quantize\_q8\_1} & 5.6 & \texttt{mul\_mat\_vec\_q\_14} & 14.9 \\
\texttt{mul\_mat\_vec\_f\_64} & 2.9 & \texttt{rms\_norm\_f32} & 5.1 & \texttt{quantize\_q8\_1} & 5.9 \\
\texttt{k\_set\_rows} & 1.9 & \texttt{mul\_mat\_vec\_f\_64} & 4.1 & \texttt{rms\_norm\_f32} & 5.3 \\
\texttt{k\_bin\_bcast\_op\_add} & 1.7 & \texttt{k\_set\_rows} & 2.5 & \texttt{mul\_mat\_vec\_f\_64} & 4.2 \\

\midrule
\multicolumn{6}{c}{Bug Fixing} \\
\cmidrule(r){1-6}
\texttt{mul\_mat\_vec\_f\_256} & 82.5 & \texttt{mul\_mat\_vec\_q\_8} & 67.2 & \texttt{mul\_mat\_vec\_q\_12} & 50.2 \\
\texttt{rms\_norm\_f32} & 3.6 & \texttt{quantize\_q8\_1} & 5.7 & \texttt{mul\_mat\_vec\_q\_14} & 15.2 \\
\texttt{k\_set\_rows} & 1.9 & \texttt{rms\_norm\_f32} & 5.1 & \texttt{quantize\_q8\_1} & 6.0 \\
\texttt{k\_bin\_bcast\_op\_add} & 1.7 & \texttt{k\_set\_rows} & 2.5 & \texttt{rms\_norm\_f32} & 5.4 \\
\texttt{mul\_mat\_f} & 1.4 & \texttt{k\_bin\_bcast\_op\_add} & 2.3 & \texttt{k\_set\_rows} & 2.7 \\

\midrule
\multicolumn{6}{c}{Docstring Generation} \\
\cmidrule(r){1-6}
\texttt{mul\_mat\_vec\_f\_256} & 82.7 & \texttt{mul\_mat\_vec\_q\_8} & 68.5 & \texttt{mul\_mat\_vec\_q\_12} & 51.2 \\
\texttt{rms\_norm\_f32} & 3.7 & \texttt{quantize\_q8\_1} & 5.8 & \texttt{mul\_mat\_vec\_q\_14} & 15.6 \\
\texttt{mul\_mat\_vec\_f\_64} & 2.9 & \texttt{rms\_norm\_f32} & 5.3 & \texttt{quantize\_q8\_1} & 6.1 \\
\texttt{k\_set\_rows} & 1.9 & \texttt{mul\_mat\_vec\_f\_64} & 4.1 & \texttt{rms\_norm\_f32} & 5.6 \\
\texttt{k\_bin\_bcast\_op\_add} & 1.7 & \texttt{k\_set\_rows} & 2.6 & \texttt{mul\_mat\_vec\_f\_64} & 4.4 \\

\midrule
\multicolumn{6}{c}{Test Generation} \\
\cmidrule(r){1-6}
\texttt{mul\_mat\_vec\_f\_256} & 82.3 & \texttt{mul\_mat\_vec\_q\_8} & 67.0 & \texttt{mul\_mat\_vec\_q\_12} & 50.2 \\
\texttt{rms\_norm\_f32} & 3.6 & \texttt{quantize\_q8\_1} & 5.7 & \texttt{mul\_mat\_vec\_q\_14} & 15.2 \\
\texttt{mul\_mat\_vec\_f\_64} & 2.3 & \texttt{rms\_norm\_f32} & 5.1 & \texttt{quantize\_q8\_1} & 6.0 \\
\texttt{k\_set\_rows} & 1.9 & \texttt{mul\_mat\_vec\_f\_64} & 3.3 & \texttt{rms\_norm\_f32} & 5.4 \\
\texttt{k\_bin\_bcast\_op\_add} & 1.7 & \texttt{k\_set\_rows} & 2.5 & \texttt{mul\_mat\_vec\_f\_64} & 3.4 \\

\bottomrule
\end{tabular}
\end{threeparttable}
\end{table*}

\textcolor{black}{\textbf{Implications.} These results can be valuable for CUDA and hardware designers, but they are also relevant to the Green AI community and AI practitioners. By showing that matrix-vector multiplication dominates inference, we provide empirical evidence that hardware accelerators should focus on optimizing these specific kernels to make LLMs more sustainable. At the same time, our findings remain actionable for developers. For example, the observation that ``quantization is not free'' highlights that lower-bit models introduce additional quantization operations, even though they consume less energy overall. This insight can help practitioners make more informed decisions when selecting model configurations.}

\section{Related Work}
As highlighted by Watson and colleagues \cite{watson2022systematic}, since 2017 DL has been used extensively in solving software engineering tasks. Moreover, the continuous evolution of larger, more capable, and computationally intensive DL models has made examining their energy consumption an increasingly important concern.
Despite the growing utilization of LLMs in software-related tasks~\cite{stack2024aihelp, github2024copilotimpact}, much of the existing research in evaluating their performance has traditionally focused on metrics such as accuracy \cite{pradel2018deepbugs, chen2024evaluating}, or task-specific metrics like pass@k for code synthesis \cite{muennighoff2023octopack, chen2021evaluating, roziere2023code, li2023starcoder} and repair rate for bug fixing \cite{zhang2024pydex}.

On the other hand, most of the existing work focuses on the energy usage of the training side \cite{anthony2020carbontracker, li2016evaluating, yarally2023uncovering, thompson2020computational, mcdonald2022great, Patternson:2022:CFM}, with few studies considering inference \cite{GeorgiouK0SZ22}. However, once the model is trained, its energy usage for inference can contribute significantly to long-term energy impact. Because the inference is often a repeated and ongoing process, performed every time a user queries the model, potentially millions of times~\cite{desislavov2021compute}. At Google~\cite{Patternson:2022:CFM}, for example, inference energy accounts for 3/5 of all the energy spent in ML tasks. 

In this section, first, we aim to outline the current state of research related to measuring energy usage during the inference phase of DL models. Luccioni et al. \cite{luccioni2022estimating} estimated the energy and carbon footprint of the BLOOM 176B model aiming to define and connect the different sources of carbon emissions involved in training and deploying. Following this, subsequent research~\cite{samsi2023words} delved explicitly into the deployment phase of the LLaMa model in 3 different sizes on 2 datasets. Their findings show that the complexity of the input dataset can affect the model's performance. They also suggest power capping as an effective tool for reducing inference energy.
In another work, Luccioni et al.~\cite{luccioni2024power} conducted a study that provides a comprehensive analysis and comparison of inference energy costs across a wide range of generative systems, including both task-specific and general-purpose models. The work utilized 88 models over 10 tasks and 30 datasets, including image generation, text completion, etc. that task-specific models are more energy efficient than multi-purpose models when performing the same task. Besides, image-related tasks are more energy-intensive than text-only tasks. Lastly, Desislavov et al.~\cite{desislavov2021compute} analyzed the energy consumption of DNNs during inference for both vision and NLP tasks. Their findings suggest that better results for DNN models are partly due to algorithmic progress, not just increased computing power. Moreover, hardware improvements can play a role in decreasing DNN energy consumption.

The rest of this section delves into studies that focus on the performance of optimization techniques for LLMs, particularly model quantization. Jin et al.~\cite{jin2024comprehensive} evaluated instruction-tuned LLMs and their quantized versions using various quantization strategies across different parameter scales. They tested these models on 7 knowledge/capacity and 3 question-answering benchmarks, focusing on the Qwen model in 7b, 14b and 72b parameters. The study reported performance metrics like accuracy and BLEU but did not consider energy consumption. They noted that quantization reduces memory usage but can slow down inference.
Tuggener et al.~\cite{tuggener2024so} reviewed 3 different quantization techniques for LlamaV2-7b. Their findings showed that quantization significantly reduces GPU memory usage but can impact text quality. They also briefly discussed the energy usage involved in fine-tuning these models.
Similarly, Rajput and Sharma~\cite{rajputbenchmarking} benchmarked and analyzed the energy efficiency of various quantization techniques.

\textcolor{black}{
As this work extends a paper submitted in 2024, several recent studies have emerged since then that specifically target inference energy costs.
Poddar et al.~\cite{poddar-etal-2025-towards} present a broad benchmarking study of inference energy costs in LLMs, examining how factors like output-token length, response time, quantization, prompt phrasing, batch size, model size, and hardware affect the total energy used during inference. Zschache and Hartwig ~\cite{zschache2025comparing} analyze several architectures and hardware to map the trade-offs between classification accuracy and energy usage, finding that higher accuracy does not always imply greater efficiency and that runtime can often serve as a proxy for energy. Adamska et al. in Green prompting ~\cite{adamska2025green} explore how prompt- and response-level characteristics (e.g., prompt length, semantics, types of tasks) directly lead to variation in energy consumption for identical tasks, showing that even without changing models, some prompting strategies yield more energy-efficient responses. Complementing these works, Fernandez et al.~\cite{fernandez-etal-2025-energy} provide a systematic analysis of how real-world inference workloads interact with efficiency interventions such as batching, decoding strategies, software frameworks, and GPU architectures. Their study demonstrates that naive FLOPs-based estimates substantially underestimate true inference energy use, and shows that applying efficiency optimizations appropriately can reduce total energy consumption by up to 73\% compared to unoptimized baselines.
\\
While recent work has begun to examine the inference energy footprint of LLMs, these studies remain largely focused on language understanding tasks such as classification, sentiment analysis, reasoning, summarization, and question answering. By contrast, our study addresses an open gap by investigating the energy cost of code understanding tasks, which differ fundamentally from natural language and introduce unique inference characteristics.
\\
In this paper, we evaluate the energy impact of 26 model families across 4 different tasks on 2 hardware configurations, while also accounting for model
accuracy.
}

\section{Threats to Validity}
A potential threat to \textit{internal validity} in this study relates to the use of prompt templates. For the sake of fair energy comparison, we used the same prompts across all models within each task, without tailoring them to individual model capabilities. This uniform approach may have resulted in differences in accuracy compared to results reported in the literature. Moreover, a few models experienced incomplete cases when using the Ollama API, where packets received were not fully completed. Although we adjusted the token limit to mitigate this issue, it persisted and may have impacted the accuracy of certain models and affected internal consistency.
Additionally, the process of extracting generated code from the LLMs' responses can also introduce a potential threat to internal validity. A few models did not include clear starting and ending markers for code sections, making it difficult to accurately detect the intended answer using regular expressions. To address this, we first employed a regular expression approach and subsequently used an open-access powerful LLM (gpt4o-mini) to extract the code from natural language explanations. Despite these efforts, this process may still negatively impact the accuracy evaluation of some models. 
\\
\textcolor{black}{In addition, each experimental configuration was executed three times to account for transient variability in system performance. While a larger number of repetitions could further improve statistical reliability, it would increase the time required to perform the experiments by an order of magnitude, which would not be feasible with our available infrastructure. Nevertheless, each trial processes a large number of queries and its execution lasts for at least a few minutes, and we are collecting observations ten times per second. This results in stable utilization patterns and reduces the influence of short-term fluctuations such as scheduling decisions and background noise.
Moreover, in our experiments, energy is computed using the arithmetic mean GPU power rather than trapezoidal integration over time, which may introduce a small approximation error. However, our validation shows that this deviation is small in practice (on average 1.8\% and maximum 4\%).
}

Another potential limitation is that our evaluation is based solely on the HumanEvalPack dataset, which includes only basic functions and may not fully represent real-world software development tasks. This constraint reduces the generalizability of our findings to more complex or varied coding scenarios, potentially impacting the \textit{external validity} of this study. 
\textcolor{black}{Additionally, our energy measurements on the A100 systems are limited merely to GPU power obtained via NVML, as CPU energy could not be collected due to lack of privileged access to RAPL interfaces. As a result, system-level energy consumption (including CPU, DRAM, and other components) is not captured for these experiments. In addition, although system-level factors such as PCIe data transfer may affect energy and runtime, they are not measured in our setup. These limitations introduce potential confounds and may lead to an underestimation of total energy consumption. However, since all experimental configurations were executed under consistent hardware settings and similar data transfer patterns, we expect their impact to be uniform across runs.
}

Furthermore, we only evaluated models using the Python programming language, which limits the diversity of real-world programming tasks covered. General models may benefit from Python’s high-level syntax, especially when generating docstrings, due to the relative ease of understanding Python functions. Including additional programming languages would improve the generalizability of the results.

Finally, the study lacks a sufficient number of comparable pairs of general-purpose models and their code-optimized counterparts. This shortcoming limits the strength of our conclusions regarding energy consumption differences between general and specialized LLMs. We mitigated this threat by analyzing a wide range of models in terms of sizes, companies that produced them, and whether they are coding-specific or non-coding-specific.

\section{Conclusion}
Our findings indicate that larger models often have higher energy consumption without proportional accuracy gains and can sometimes be outperformed by smaller models, suggesting that model size alone is not a reliable predictor of accuracy. We found that a model's energy usage varies significantly across software development tasks, highlighting the importance of selecting models fine-tuned to specific applications to manage energy consumption effectively and maximize efficiency.
Although energy usage varies across tasks, the energy consumption per generated token remains consistent for a given model. Moreover, the strong correlation between total energy consumption and architectural characteristics, particularly active parameter count, suggests that an LLM's architecture can serve as a useful indicator of its energy efficiency, given that the average size of the outputs can be anticipated. \textcolor{black}{Additionally, coding-specific models achieve higher accuracy when energy is not considered, while they rarely appear on the Pareto frontier once energy consumption is taken into account. The results suggest that current task-specific models are primarily optimized for accuracy rather than energy efficiency. It emphasizes the need for future coding models that balance both accuracy and energy efficiency across software engineering tasks.
Finally, our regression analysis revealed that active parameter count, output token count, and quantization level are the primary predictors of inference energy consumption, with statistically significant effects across all tasks. \textcolor{black}{
In addition, the interaction between output tokens and active parameter count is consistently significant, indicating that the cost of generating tokens increases with model size. Together, these factors explain 89\%, 92\%, 92\%, and 94\% of the variance in total energy use for \code, \bug, \doc, and \test, respectively, indicating a strong model fit. In contrast, prompt size has a negligible impact on energy usage once model size, output length, and quantization are controlled for.
}}
\textcolor{black}{Future work will extend this study by exploring a wider range of prompt sizes to better assess the impact of input length on energy consumption, and by incorporating additional programming languages and more diverse benchmarks, including real-world coding tasks, to evaluate model behavior under realistic conditions and to improve the generalizability of our findings.}
\label{sec:conc}

\section*{Data Availability}
The replication package for this study, along with the appendices, is publicly available~\cite{Alizadeh2026_replication}.

\section*{Declarations}

\textbf{Conflicts of interest.} The authors declare that they have no conflicts of interest.
\\
\textbf{Clinical trial number.} Not applicable.
\\
\textbf{Ethical approval.} This study does not involve human participants or animals.
\\
\textbf{Informed consent.} Not applicable.

\bibliographystyle{spbasic}
\bibliography{references.bib}
\newpage
\section{Appendices}
\begin{table*}[!htp]\centering
\begin{threeparttable}
\caption{GPU A100 – Energy and Efficiency results for omitted models (complementary to Table~\ref{tab:a100_res_ee}).}\label{tab:a100_res_app_ee}
\scriptsize
\begin{tabular}{l@{\hskip 2pt}rrrr||rrrr}\toprule
\multirow{2}{*}{Model name} & \multicolumn{4}{c}{Total energy (Wh)} & \multicolumn{4}{c}{Efficiency (tokens/J)} \\
\cmidrule(lr){2-5}\cmidrule(lr){6-9}
 & CG & BF & DG & TG & CG & BF & DG & TG \\\midrule
\cellcolor[HTML]{E8E8E8}Codegemma:7b-fp16\tnote{*}  &\cellcolor[HTML]{E8E8E8}29.17 &\cellcolor[HTML]{E8E8E8}45.89 &\cellcolor[HTML]{E8E8E8}24.88 &\cellcolor[HTML]{E8E8E8}115.09 &\cellcolor[HTML]{E8E8E8}0.28 &\cellcolor[HTML]{E8E8E8}0.27 &\cellcolor[HTML]{E8E8E8}0.28 &\cellcolor[HTML]{E8E8E8}0.27 \\
Codegemma:7b-q8 &24.08 &36.75 &19.60 &89.05 &0.34 &0.34 &0.35 &0.35 \\
\cellcolor[HTML]{E8E8E8}Codegemma:7b-q4 &\cellcolor[HTML]{E8E8E8}20.56 &\cellcolor[HTML]{E8E8E8}28.56 &\cellcolor[HTML]{E8E8E8}15.70 &\cellcolor[HTML]{E8E8E8}75.03 &\cellcolor[HTML]{E8E8E8}0.43 &\cellcolor[HTML]{E8E8E8}0.42 &\cellcolor[HTML]{E8E8E8}0.44 &\cellcolor[HTML]{E8E8E8}0.43 \\
Deepseek-coder:1.3b-fp16\tnote{*}  &10.81 &20.12 &5.91 &14.06 &1.21 &1.20 &1.25 &1.20 \\
\cellcolor[HTML]{E8E8E8}Deepseek-coder:1.3b-q8 &\cellcolor[HTML]{E8E8E8}8.03 &\cellcolor[HTML]{E8E8E8}14.59 &\cellcolor[HTML]{E8E8E8}4.04 &\cellcolor[HTML]{E8E8E8}11.16 &\cellcolor[HTML]{E8E8E8}1.63 &\cellcolor[HTML]{E8E8E8}1.60 &\cellcolor[HTML]{E8E8E8}1.75 &\cellcolor[HTML]{E8E8E8}1.62 \\
Deepseek-coder:1.3b-q4 &6.16 &\textbf{11.40} &\textbf{2.96} &\textbf{7.95} &2.07 &2.00 &2.16 &2.04 \\
\cellcolor[HTML]{E8E8E8}Deepseek-llm:7b-fp16 &\cellcolor[HTML]{E8E8E8}13.98 &\cellcolor[HTML]{E8E8E8}28.83 &\cellcolor[HTML]{E8E8E8}22.03 &\cellcolor[HTML]{E8E8E8}23.94 &\cellcolor[HTML]{E8E8E8}0.36 &\cellcolor[HTML]{E8E8E8}0.35 &\cellcolor[HTML]{E8E8E8}0.36 &\cellcolor[HTML]{E8E8E8}0.35 \\
Deepseek-llm:7b-q8 &10.00 &20.90 &15.70 &16.73 &0.50 &0.50 &0.51 &0.50 \\
\cellcolor[HTML]{E8E8E8}Deepseek-llm:7b-q4 &\cellcolor[HTML]{E8E8E8}10.57 &\cellcolor[HTML]{E8E8E8}19.28 &\cellcolor[HTML]{E8E8E8}13.96 &\cellcolor[HTML]{E8E8E8}21.21 &\cellcolor[HTML]{E8E8E8}0.53 &\cellcolor[HTML]{E8E8E8}0.52 &\cellcolor[HTML]{E8E8E8}0.55 &\cellcolor[HTML]{E8E8E8}0.53 \\
Llama2:13b-fp16 &13.09 &61.76 &33.19 &94.75 &0.20 &0.19 &0.20 &0.20 \\
\cellcolor[HTML]{E8E8E8}Llama2:13b-q8 &\cellcolor[HTML]{E8E8E8}10.05 &\cellcolor[HTML]{E8E8E8}46.08 &\cellcolor[HTML]{E8E8E8}23.37 &\cellcolor[HTML]{E8E8E8}67.02 &\cellcolor[HTML]{E8E8E8}0.27 &\cellcolor[HTML]{E8E8E8}0.28 &\cellcolor[HTML]{E8E8E8}0.28 &\cellcolor[HTML]{E8E8E8}0.28 \\
Llama2:13b-q4 &11.03 &31.54 &19.90 &46.80 &0.32 &0.32 &0.34 &0.33 \\
\cellcolor[HTML]{E8E8E8}Llama2:7b-fp16 &\cellcolor[HTML]{E8E8E8}7.48 &\cellcolor[HTML]{E8E8E8}21.51 &\cellcolor[HTML]{E8E8E8}18.55 &\cellcolor[HTML]{E8E8E8}60.72 &\cellcolor[HTML]{E8E8E8}0.36 &\cellcolor[HTML]{E8E8E8}0.35 &\cellcolor[HTML]{E8E8E8}0.36 &\cellcolor[HTML]{E8E8E8}0.36 \\
Llama2:7b-q8 &\textbf{5.61} &15.45 &12.67 &41.92 &0.50 &0.49 &0.52 &0.51 \\
\cellcolor[HTML]{E8E8E8}Llama2:7b-q4 &\cellcolor[HTML]{E8E8E8}5.92 &\cellcolor[HTML]{E8E8E8}17.56 &\cellcolor[HTML]{E8E8E8}11.03 &\cellcolor[HTML]{E8E8E8}30.43 &\cellcolor[HTML]{E8E8E8}0.53 &\cellcolor[HTML]{E8E8E8}0.53 &\cellcolor[HTML]{E8E8E8}0.56 &\cellcolor[HTML]{E8E8E8}0.55 \\
Mistral:7b-fp16 &12.41 &36.20 &17.58 &59.73 &0.34 &0.34 &0.35 &0.34 \\
\cellcolor[HTML]{E8E8E8}Mistral:7b-q8 &\cellcolor[HTML]{E8E8E8}9.07 &\cellcolor[HTML]{E8E8E8}25.85 &\cellcolor[HTML]{E8E8E8}12.37 &\cellcolor[HTML]{E8E8E8}41.08 &\cellcolor[HTML]{E8E8E8}0.49 &\cellcolor[HTML]{E8E8E8}0.49 &\cellcolor[HTML]{E8E8E8}0.50 &\cellcolor[HTML]{E8E8E8}0.50 \\
Mistral:7b-q4 &7.97 &24.49 &11.71 &37.96 &0.52 &0.53 &0.54 &0.54 \\
\bottomrule
\end{tabular}
\begin{tablenotes}
    \footnotesize
    \item[*] Code-specific model families.
\end{tablenotes}
\end{threeparttable}
\end{table*}

\begin{table*}[!htp]\centering
\begin{threeparttable}
\caption{GPU A100 – Accuracy results for omitted models (complementary to Table~\ref{tab:a100_res_ac}). SC and BC denote statement and branch coverage.}\label{tab:a100_res_app_ac}
\scriptsize
\begin{tabular}{lrrr|rrr}\toprule
\multirow{2}{*}{Model name} &\multicolumn{4}{c}{Accuracy\%}  & \multicolumn{1}{c}{SC\%} &\multicolumn{1}{c}{BC\%} \\\cmidrule(lr){2-5}\cmidrule(lr){6-7}
&CG &BF &DG &TG & & \\\midrule
\cellcolor[HTML]{E8E8E8}Codegemma:7b-fp16\tnote{*}  &\cellcolor[HTML]{E8E8E8}\textbf{52.44} &\cellcolor[HTML]{E8E8E8}21.95 &\cellcolor[HTML]{E8E8E8}54.88 &\cellcolor[HTML]{E8E8E8}17.07 &\cellcolor[HTML]{E8E8E8}60.55 &\cellcolor[HTML]{E8E8E8}77.58 \\
Codegemma:7b-q8 &\textbf{52.44} &22.56 &57.32 &18.90 &64.85 &79.31 \\
\cellcolor[HTML]{E8E8E8}Codegemma:7b-q4 &\cellcolor[HTML]{E8E8E8}49.40 &\cellcolor[HTML]{E8E8E8}22.56 &\cellcolor[HTML]{E8E8E8}53.66 &\cellcolor[HTML]{E8E8E8}18.29 &\cellcolor[HTML]{E8E8E8}60.39 &\cellcolor[HTML]{E8E8E8}77.86 \\
Deepseek-coder:1.3b-fp16\tnote{*}  &7.32 &7.93 &43.29 &27.44 &72.82 &50.10 \\
\cellcolor[HTML]{E8E8E8}Deepseek-coder:1.3b-q8 &\cellcolor[HTML]{E8E8E8}7.32 &\cellcolor[HTML]{E8E8E8}7.93 &\cellcolor[HTML]{E8E8E8}44.51 &\cellcolor[HTML]{E8E8E8}28.05 &\cellcolor[HTML]{E8E8E8}76.92 &\cellcolor[HTML]{E8E8E8}57.11 \\
Deepseek-coder:1.3b-q4 &3.05 &6.71 &42.07 &29.88 &76.19 &55.17 \\
\cellcolor[HTML]{E8E8E8}Deepseek-llm:7b-fp16 &\cellcolor[HTML]{E8E8E8}28.05 &\cellcolor[HTML]{E8E8E8}13.41 &\cellcolor[HTML]{E8E8E8}59.76 &\cellcolor[HTML]{E8E8E8}\textbf{43.90} &\cellcolor[HTML]{E8E8E8}77.31 &\cellcolor[HTML]{E8E8E8}63.28 \\
Deepseek-llm:7b-q8 &31.10 &10.98 &\textbf{61.59} &41.46 &78.41 &65.34 \\
\cellcolor[HTML]{E8E8E8}Deepseek-llm:7b-q4 &\cellcolor[HTML]{E8E8E8}25.00 &\cellcolor[HTML]{E8E8E8}11.59 &\cellcolor[HTML]{E8E8E8}59.15 &\cellcolor[HTML]{E8E8E8}35.98 &\cellcolor[HTML]{E8E8E8}66.64 &\cellcolor[HTML]{E8E8E8}57.81 \\
Llama2:13b-fp16 &13.41 &10.37 &41.46 &35.37 &81.60 &75.36 \\
\cellcolor[HTML]{E8E8E8}Llama2:13b-q8 &\cellcolor[HTML]{E8E8E8}13.41 &\cellcolor[HTML]{E8E8E8}9.76 &\cellcolor[HTML]{E8E8E8}33.54 &\cellcolor[HTML]{E8E8E8}32.32 &\cellcolor[HTML]{E8E8E8}81.85 &\cellcolor[HTML]{E8E8E8}75.07 \\
Llama2:13b-q4 &10.98 &11.59 &39.02 &30.49 &76.27 &60.93 \\
\cellcolor[HTML]{E8E8E8}Llama2:7b-fp16 &\cellcolor[HTML]{E8E8E8}6.10 &\cellcolor[HTML]{E8E8E8}6.71 &\cellcolor[HTML]{E8E8E8}39.02 &\cellcolor[HTML]{E8E8E8}20.73 &\cellcolor[HTML]{E8E8E8}72.74 &\cellcolor[HTML]{E8E8E8}71.94 \\
Llama2:7b-q8 &7.32 &6.10 &38.41 &21.34 &72.75 &69.75 \\
\cellcolor[HTML]{E8E8E8}Llama2:7b-q4 &\cellcolor[HTML]{E8E8E8}6.10 &\cellcolor[HTML]{E8E8E8}4.88 &\cellcolor[HTML]{E8E8E8}36.59 &\cellcolor[HTML]{E8E8E8}40.24 &\cellcolor[HTML]{E8E8E8}75.67 &\cellcolor[HTML]{E8E8E8}64.99 \\
Mistral:7b-fp16 &7.93 &\textbf{27.44} &51.83 &31.71 &72.23 &69.17 \\
\cellcolor[HTML]{E8E8E8}Mistral:7b-q8 &\cellcolor[HTML]{E8E8E8}3.66 &\cellcolor[HTML]{E8E8E8}25.61 &\cellcolor[HTML]{E8E8E8}53.05 &\cellcolor[HTML]{E8E8E8}28.66 &\cellcolor[HTML]{E8E8E8}72.54 &\cellcolor[HTML]{E8E8E8}69.59 \\
Mistral:7b-q4 &13.41 &24.39 &53.66 &24.39 &63.38 &66.16 \\
\bottomrule
\end{tabular}
\begin{tablenotes}
    \footnotesize
    \item[*] Code-specific model families.
\end{tablenotes}
\end{threeparttable}
\end{table*}
\begin{table}[!ht]
\centering
\caption{\revision{Regression results for \textbf{energy consumption} in the comparison of general-purpose and coding-specific models. Task dummies are included with \doc~as the reference category. Cluster-Robust SEs are used (clustered by family). Number of observations is 312 and $R^2$ = 0.939}}
\label{tab:reg_energy_coding_general}
\begin{tabular}{lccc}
\toprule
Variable & Coefficient ($\beta$) & SE & p-value \\
\midrule
coding\_model & -0.62 & 0.82 & 0.454 \\
coding\_model $\times$ code & 0.13 & 0.28 & 0.641 \\
coding\_model $\times$ bug & 0.01 & 0.66 & 0.984 \\
coding\_model $\times$ test & 0.39 & 0.87 & 0.649 \\
\bottomrule
\end{tabular}
\end{table}

\begin{table}[!ht]
\centering
\caption{\revision{Regression results for \textbf{accuracy} in the comparison of general-purpose and coding-specific models. Task dummies are included with \doc~as the reference category. Cluster-Robust SEs are used (clustered by family). Number of observations is 312 and $R^2$ = 0.551}}
\label{tab:reg_acc_coding_general}
\begin{tabular}{lccc}
\toprule
Variable & Coefficient ($\beta$) & SE & p-value \\
\midrule
coding\_model & 1.48 & 4.85 & 0.760 \\
coding\_model $\times$ code & 13.14 & 5.99 & 0.028 \\
coding\_model $\times$ bug & 9.03 & 4.55 & 0.047 \\
coding\_model $\times$ test & 18.78 & 8.06 & 0.020 \\
\bottomrule
\end{tabular}
\end{table}

\begin{table*}[!htp]\centering
\begin{threeparttable}
\caption{GPU A100 – Distribution of kernel execution time (\%) across quantization levels (FP16, Q8, Q4) for different software engineering tasks when running \texttt{Qwen2.5-coder:7B}.}
\label{tab:nsys_qwen}
\scriptsize
\begin{tabular}{l r | l r | l r}
\toprule
\multicolumn{2}{c|}{FP16} & \multicolumn{2}{c|}{Q8} & \multicolumn{2}{c}{Q4} \\
Kernel & (\%) & Kernel & (\%) & Kernel & (\%) \\
\midrule

\multicolumn{6}{c}{Code Generation} \\
\cmidrule(r){1-6}
\texttt{mul\_mat\_vec\_f\_256} & 83.4 & \texttt{mul\_mat\_vec\_q\_8} & 72.6 & \texttt{mul\_mat\_vec\_q\_12} & 47.5 \\
\texttt{k\_bin\_bcast\_op\_add} & 3.3 & \texttt{k\_bin\_bcast\_op\_add} & 4.1 & \texttt{mul\_mat\_vec\_q\_14} & 18.5 \\
\texttt{rms\_norm\_f32} & 3.0 & \texttt{quantize\_q8\_1} & 4.0 & \texttt{quantize\_q8\_1} & 4.9 \\
\texttt{mul\_mat\_vec\_f\_64} & 1.9 & \texttt{rms\_norm\_f32} & 3.7 & \texttt{k\_bin\_bcast\_op\_add} & 4.9 \\
\texttt{k\_set\_rows} & 1.4 & \texttt{mul\_mat\_vec\_f\_64} & 2.4 & \texttt{rms\_norm\_f32} & 4.5 \\

\midrule
\multicolumn{6}{c}{Bug Fixing} \\
\cmidrule(r){1-6}
\texttt{mul\_mat\_vec\_f\_256} & 84.4 & \texttt{mul\_mat\_vec\_q\_8} & 74.0 & \texttt{mul\_mat\_vec\_q\_12} & 49.0 \\
\texttt{k\_bin\_bcast\_op\_add} & 3.3 & \texttt{k\_bin\_bcast\_op\_add} & 4.1 & \texttt{mul\_mat\_vec\_q\_14} & 19.0 \\
\texttt{rms\_norm\_f32} & 3.0 & \texttt{quantize\_q8\_1} & 4.1 & \texttt{quantize\_q8\_1} & 5.1 \\
\texttt{k\_set\_rows} & 1.4 & \texttt{rms\_norm\_f32} & 3.7 & \texttt{k\_bin\_bcast\_op\_add} & 5.0 \\
\texttt{mul\_mat\_vec\_f\_64} & 1.4 & \texttt{k\_set\_rows} & 1.7 & \texttt{rms\_norm\_f32} & 4.6 \\

\midrule
\multicolumn{6}{c}{Docstring Generation} \\
\cmidrule(r){1-6}
\texttt{mul\_mat\_vec\_f\_256} & 85.0 & \texttt{mul\_mat\_vec\_q\_8} & 75.8 & \texttt{mul\_mat\_vec\_q\_12} & 50.3 \\
\texttt{k\_bin\_bcast\_op\_add} & 3.3 & \texttt{k\_bin\_bcast\_op\_add} & 4.2 & \texttt{mul\_mat\_vec\_q\_14} & 19.5 \\
\texttt{rms\_norm\_f32} & 3.0 & \texttt{quantize\_q8\_1} & 4.2 & \texttt{quantize\_q8\_1} & 5.2 \\
\texttt{mul\_mat\_vec\_f\_64} & 1.9 & \texttt{rms\_norm\_f32} & 3.8 & \texttt{k\_bin\_bcast\_op\_add} & 5.1 \\
\texttt{k\_set\_rows} & 1.4 & \texttt{mul\_mat\_vec\_f\_64} & 2.4 & \texttt{rms\_norm\_f32} & 4.8 \\

\midrule
\multicolumn{6}{c}{Test Generation} \\
\cmidrule(r){1-6}
\texttt{mul\_mat\_vec\_f\_256} & 84.3 & \texttt{mul\_mat\_vec\_q\_8} & 74.1 & \texttt{mul\_mat\_vec\_q\_12} & 48.9 \\
\texttt{k\_bin\_bcast\_op\_add} & 3.3 & \texttt{k\_bin\_bcast\_op\_add} & 4.1 & \texttt{mul\_mat\_vec\_q\_14} & 19.0 \\
\texttt{rms\_norm\_f32} & 3.0 & \texttt{quantize\_q8\_1} & 4.1 & \texttt{quantize\_q8\_1} & 5.1 \\
\texttt{mul\_mat\_vec\_f\_64} & 1.8 & \texttt{rms\_norm\_f32} & 3.7 & \texttt{k\_bin\_bcast\_op\_add} & 5.0 \\
\texttt{k\_set\_rows} & 1.4 & \texttt{mul\_mat\_vec\_f\_64} & 2.3 & \texttt{rms\_norm\_f32} & 4.6 \\

\bottomrule
\end{tabular}
\end{threeparttable}
\end{table*}

\begin{table*}[!htp]\centering
\begin{threeparttable}
\caption{GPU A100 – Distribution of kernel execution time (\%) across quantization levels (FP16, Q8, Q4) for different software engineering tasks when running \texttt{Starcoder2:15b}.}
\label{tab:nsys_starcoder}
\scriptsize
\begin{tabular}{l r | l r | l r}
\toprule
\multicolumn{2}{c|}{FP16} & \multicolumn{2}{c|}{Q8} & \multicolumn{2}{c}{Q4} \\
Kernel & (\%) & Kernel & (\%) & Kernel & (\%) \\
\midrule

\multicolumn{6}{c}{Code Generation} \\
\cmidrule(r){1-6}
\texttt{mul\_mat\_vec\_f\_256} & 83.4 & \texttt{mul\_mat\_vec\_q\_8} & 71.7 & \texttt{mul\_mat\_vec\_q\_12} & 46.2 \\
\texttt{k\_bin\_bcast\_op\_add} & 5.1 & \texttt{k\_bin\_bcast\_op\_add} & 6.9 & \texttt{mul\_mat\_vec\_q\_14} & 19.6 \\
\texttt{mul\_mat\_vec\_f\_64} & 2.2 & \texttt{mul\_mat\_vec\_f\_64} & 3.0 & \texttt{k\_bin\_bcast\_op\_add} & 8.3 \\
\texttt{norm\_f32} & 1.5 & \texttt{quantize\_q8\_1} & 2.9 & \texttt{mul\_mat\_vec\_f\_64} & 3.6 \\
\texttt{k\_bin\_bcast\_op\_mul} & 1.0 & \texttt{norm\_f32} & 2.1 & \texttt{quantize\_q8\_1} & 3.4 \\

\midrule
\multicolumn{6}{c}{Bug Fixing} \\
\cmidrule(r){1-6}
\texttt{mul\_mat\_vec\_f\_256} & 82.0 & \texttt{mul\_mat\_vec\_q\_8} & 68.7 & \texttt{mul\_mat\_vec\_q\_12} & 44.1 \\
\texttt{k\_bin\_bcast\_op\_add} & 5.1 & \texttt{k\_bin\_bcast\_op\_add} & 6.8 & \texttt{mul\_mat\_vec\_q\_14} & 18.6 \\
\texttt{mul\_mat\_vec\_f\_64} & 1.8 & \texttt{mul\_mat\_q\_8} & 3.4 & \texttt{k\_bin\_bcast\_op\_add} & 8.1 \\
\texttt{norm\_f32} & 1.5 & \texttt{quantize\_q8\_1} & 2.8 & \texttt{mul\_mat\_q\_12} & 3.3 \\
\texttt{k\_bin\_bcast\_op\_mul} & 1.0 & \texttt{mul\_mat\_vec\_f\_64} & 2.3 & \texttt{quantize\_q8\_1} & 3.2 \\

\midrule
\multicolumn{6}{c}{Docstring Generation} \\
\cmidrule(r){1-6}
\texttt{mul\_mat\_vec\_f\_256} & 84.6 & \texttt{mul\_mat\_vec\_q\_8} & 74.4 & \texttt{mul\_mat\_vec\_q\_12} & 48.5 \\
\texttt{k\_bin\_bcast\_op\_add} & 5.0 & \texttt{k\_bin\_bcast\_op\_add} & 7.1 & \texttt{mul\_mat\_vec\_q\_14} & 20.5 \\
\texttt{mul\_mat\_vec\_f\_64} & 2.2 & \texttt{mul\_mat\_vec\_f\_64} & 3.0 & \texttt{k\_bin\_bcast\_op\_add} & 8.5 \\
\texttt{norm\_f32} & 1.5 & \texttt{quantize\_q8\_1} & 3.0 & \texttt{mul\_mat\_vec\_f\_64} & 3.7 \\
\texttt{k\_bin\_bcast\_op\_mul} & 1.0 & \texttt{norm\_f32} & 2.2 & \texttt{quantize\_q8\_1} & 3.5 \\

\midrule
\multicolumn{6}{c}{Test Generation} \\
\cmidrule(r){1-6}
\texttt{mul\_mat\_vec\_f\_256} & 84.6 & \texttt{mul\_mat\_vec\_q\_8} & 73.8 & \texttt{mul\_mat\_vec\_q\_12} & 47.7 \\
\texttt{k\_bin\_bcast\_op\_add} & 5.0 & \texttt{k\_bin\_bcast\_op\_add} & 7.0 & \texttt{mul\_mat\_vec\_q\_14} & 20.1 \\
\texttt{mul\_mat\_vec\_f\_64} & 1.8 & \texttt{quantize\_q8\_1} & 3.0 & \texttt{k\_bin\_bcast\_op\_add} & 8.4 \\
\texttt{norm\_f32} & 1.5 & \texttt{mul\_mat\_vec\_f\_64} & 2.4 & \texttt{quantize\_q8\_1} & 3.5 \\
\texttt{k\_bin\_bcast\_op\_mul} & 1.0 & \texttt{norm\_f32} & 2.2 & \texttt{mul\_mat\_vec\_f\_64} & 3.0 \\

\bottomrule
\end{tabular}
\end{threeparttable}
\end{table*}

\begin{figure}[!ht]
  \centering
  \begin{subfigure}[b]{\textwidth}
    \includegraphics[width=\textwidth]{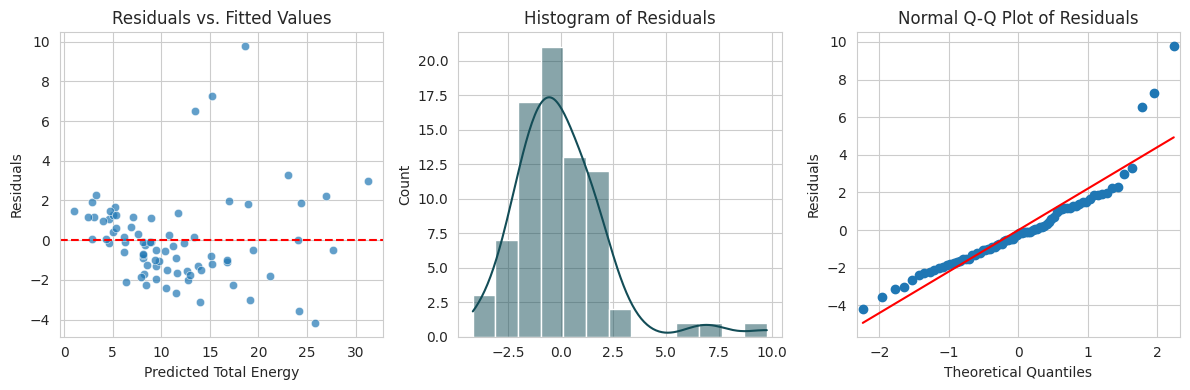}
    \caption{Code Generation}
  \end{subfigure}

  \vspace{0.3cm}

  \begin{subfigure}[b]{\textwidth}
    \includegraphics[width=\textwidth]{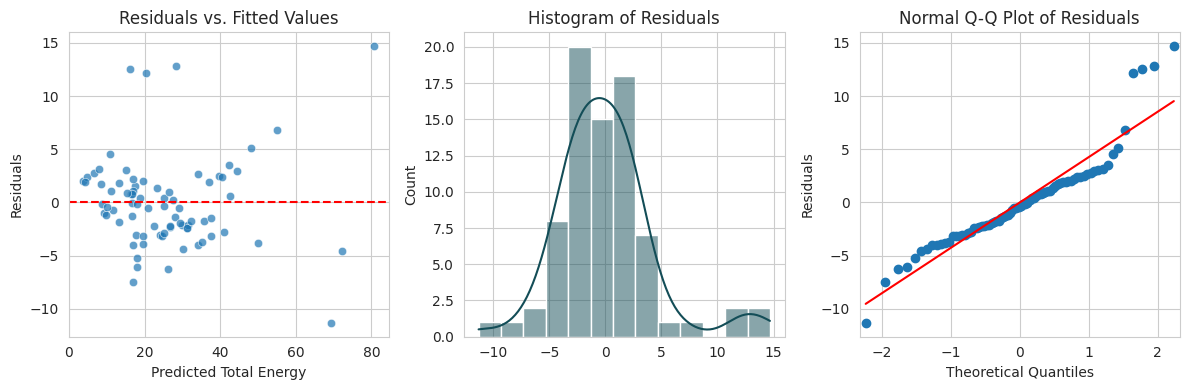}
    \caption{Bug Fixing}
  \end{subfigure}

  \vspace{0.3cm}

  \begin{subfigure}[b]{\textwidth}
    \includegraphics[width=\textwidth]{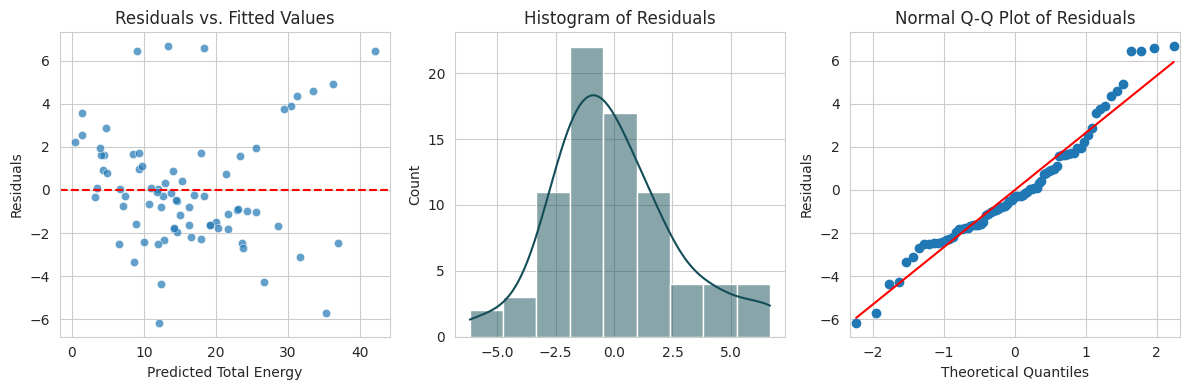}
    \caption{Docstring Generation}
  \end{subfigure}

  \vspace{0.3cm}

  \begin{subfigure}[b]{\textwidth}
    \includegraphics[width=\textwidth]{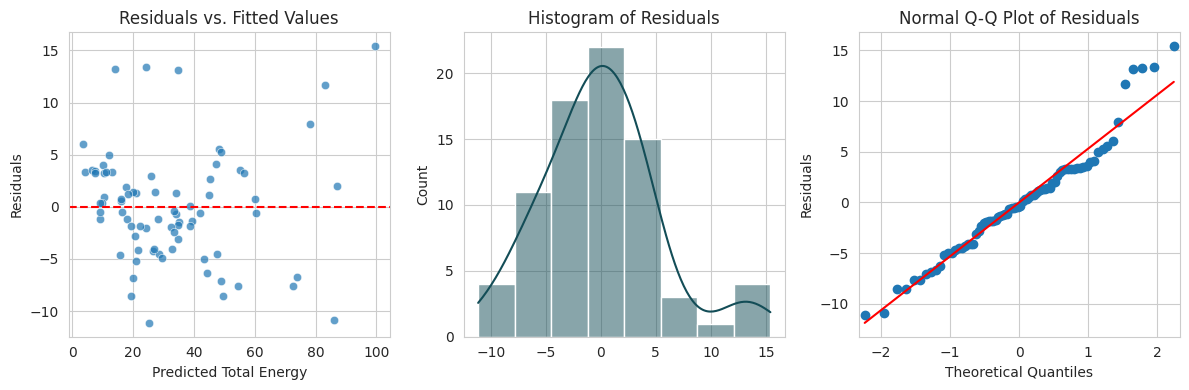}
    \caption{Test Generation}
  \end{subfigure}

  \caption{Diagnostic plots of linear regression assumptions for different tasks. The residuals show no systematic pattern (left), approximate normal distribution (middle), and reasonable alignment with theoretical quantiles (right), indicating that linearity and normality assumptions are reasonably satisfied.}
\label{fig:linearity_diagnosis}
\end{figure}
\label{sec:app}

\end{document}